\documentclass[12pt,a4paper]{iopart}
\usepackage{epsfig}  
\begin{document}

\article[Kaon-Deuteron Scattering at Low Energies]
{TOPICAL REVIEW}{Kaon-Deuteron Scattering at Low Energies}
\author{A~Sibirtsev\dag\, J~Haidenbauer\ddag\,
S~Krewald\ddag\ and Ulf-G~Mei{\ss}ner\dag\ddag }

\address{\dag\ Helmholtz-Institut f\"ur Strahlen- und Kernphysik
(Theorie), Universit\"at Bonn, Nu\ss allee 14-16, D-53115 Bonn, Germany,}

\address{\ddag\ Institut f\"ur Kernphysik, Forschungszentrum J\"ulich,
D-52425 J\"ulich, Germany}

\begin{abstract}

We review the experimental 
information on the $K^+d$ reaction for $K^+$-meson momenta below
800~MeV/c. The data are analysed within the single scattering impulse
approximation -- utilizing  the J\"ulich kaon-nucleon model --
that allows to take 
into account effects due to the Fermi motion of the nucleons in the
deuteron and the final three-body kinematics for the break-up 
and charge exchange reaction. We discuss the consistency between 
the data available for the $K^+d{\to}K^+np$, $K^+d{\to}K^0pp$ and
$K^+d{\to}K^+d$ reactions and the calculations based on the spectator 
model formalism. 
\end{abstract}



\section{Introduction}

Recently the kaon-nucleon ($K^+N$) interaction has attracted considerable 
interest because of the possible existence of the $\Theta^+$(1540)
pentaquark. The $K^+N$ system constitutes the only open hadronic decay 
channel for that resonance and, therefore, can be used to impose constraints 
on its width. The relevant isospin $I{=}0$ $K^+N$ channel can be accessed 
experimentally only via kaon-deuteron ($K^+d$) scattering and, therefore, 
pertinent analyses were performed through a direct 
inspection of data for the total $K^+d$ cross
section~\cite{Gibbs,Nussinov} and the charge exchange reaction $K^+d{\to}K^0pp$
\cite{Cahn,Sibirtsev1} in the relevant low-energy region. 
An examination~\cite{Haidenbauer} of data on the $I{=}0$ total 
$KN$ cross section was based on the isospin decomposition~\cite{Carroll} of
the $K^+d$ total cross section. A reexamination of the $KN$ partial wave (PW) 
analysis performed by Arndt et al.~\cite{Arndt} in the light of the 
$\Theta^+$(1540) followed the same procedure as their earlier extraction of
the $I{=}0$ $K^+N$ amplitudes from the $K^+d$ data. 

In the present paper we provide an overview of the experimental 
information on the $K^+d$ reaction.
We concentrate on $K^+$-meson momenta below 800~MeV/c, say, because
here the elementary $KN$ reaction is 
predominantly elastic. Judging from the available data the contribution
of inelastic channels ($KN \to KN\pi$ etc.) to the cross sections is 
still only about 10 \% at the upper end of this momentum range.
Our main goal is to investigate whether the available $K^+d$ data are 
consistent with each other and whether they can be consistently 
described using a $KN$ model that reproduces the results of 
up-to-date partial wave analyses. 
In particular, we utilize here the J\"ulich $KN$ model I from
Ref. \cite{Juel2}. The calculation of the $K^+d$ observables is carried out 
in the single scattering impulse approximation. It allows to take 
into account effects due to the Fermi motion of the nucleons within the
deuteron and the final three-body kinematics for the break-up reaction. 
The same formalism was applied by us previously in the analysis of 
the reaction $K^+d{\to}K^0pp$. In the present study we concentrate
on the other $K^+d$ channels where data are available, i.e. the
reactions $K^+d{\to}K^+np$ and $K^+d{\to}K^+d$.
Note that this formalism forms also the basis of practically all
$KN$ partial wave analyses \cite{Martin,Watts,Nakajima,Hashimoto1,Hyslop}. 

In the literature one can find only a small number of $K^+d$ studies 
which were performed within a three-body (Faddeev) framework 
\cite{Hetherington,Sanudo,Garcilazo} or where at least corrections from 
multiple scattering were taken into account 
\cite{Gourdin,Andrade,Hashimoto,Hashimoto1,Sanudo1}.
Furthermore, all those works concentrated on specific reactions and 
thus only on a rather limited set of the available $K^+d$ data. 
With regard to $K^+d$ coherent scattering the earliest Faddeev-type 
calculations~\cite{Hetherington}, performed for kaon momenta of 
110-230 MeV/c, suggested that in this momentum range multiple scattering 
corrections are of the order of 10-20\%. 
This was later on confirmed by relativistic 
Faddeev calculations presented by Garcilazo~\cite{Garcilazo} which
covered the momentum range up to 1.5 GeV/c. His results also demonstrate
that 
the total and the elastic $K^+d$ cross section obtained from the full
Faddeev calculation and from the impulse approximation practically
coincide for kaon momenta of 400 MeV/c onwards, say. 

Low-energy charge-exchange $K^+d$ scattering was studied by Sa\~nudo 
\cite{Sanudo1} within a multiple scattering expansion. He concluded
that double-scattering terms due to the $KN$ system and higher
ones due to $KN$ and $NN$ systems had no observable effect on the
charge-exchange $K^+d$ cross section above 252 MeV/c. But even
the effect of the $NN$ final-state interaction turned out to be
rather small \cite{Sanudo1}. 

In view of those results we anticipate that the impulse approximation
that we use in our calculation should work rather well,
in particular for momenta above 400 MeV/c where the bulk of the 
$K^+d$ data are. Still effects from multiple scattering are
expected to play a role for specific reaction kinematics. But
the main question is, of course, whether those are manifest in 
the presently available empirical information on $K^+d$ scattering, 
specifically, given the large error bars of the data.
In any case, 
possible discrepancies between our calculation and experimental
observables could be a signal for such effects. 

In the present work we aim primarily for a direct comparision of our 
calculation with measured $K^+d$ observables. But we consider also
published $K^+n$ data, which have been inferred from the reaction
$K^+d \to K^+np$. The extraction of those 
$K^+n$ scattering data from the deuteron reaction is based on the 
so-called spectator model, i.e. the single scattering impulse 
approximation. Thereby, it is 
assumed that the proton of the deuteron is the spectator and its 
only role in the $K^+n$ interaction is due to the Fermi motion of the 
bound neutron. This is, in principle, a reliable method
\cite{Meyer,Duncan,Calen,Zlomanczuk,Moskal} as long as one
measures the momentum distribution of the proton and one takes only 
those events which fulfil the so-called spectator condition, i.e.
with the momentum of the proton being less than the momentum of the
neutron ($p_p{<}p_n$). However, in practice often  
the spectator proton and the final neutron were not even identified  
in the corresponding $K^+d$ experiments. 

Another difficulty arises in case of the
$I{=}0$ total $K^+N$ cross section. It 
was extracted~\cite{Carroll,Bowen1,Bowen2,Bugg1,Cool,Giacomelli3} 
from the total $K^+d$ cross section utilizing the
Glauber formalism, which originally was proposed
for high energies~\cite{Glauber,Franco}. The shadowing corrections, which
appear in these analyses, are based on substantial contributions from higher 
partial waves in the elementary scattering amplitude, i.e. it is assumed 
implicitly that the scattering amplitude dominates at forward direction. 
That is clearly not the case for $KN$ scattering at low momenta.

Our paper is organized as follows. The J\"ulich model for the $KN$
interaction is described in Sect.~2. In Sect.~3 we briefly give the 
formalism used for the $K^+d$ charge exchange and break-up reactions and then we
analyse the $K^+d{\to}K^+np$ and $K^+d{\to}K^0pp$ data. 
The formalism for coherent $K^+$
scattering is given in Sect.~4 together with an analysis of the
$K^+d{\to}K^+d$ data. The total $K^+d$ cross section is discussed in
Sect.~5. The paper ends with summary.

\section{The J\"ulich meson-exchange kaon-nucleon model.}

The J\"ulich model of the $KN$ interaction has 
been described in detail in Refs.~\cite{Juel2,Juel1,Juel3,Juel4}.
Thus, we summarize here only the main features.
The J\"ulich meson-exchange model of the $KN$ interaction was 
constructed along the lines of the Bonn $NN$ model~\cite{MHE} and its 
extension to the hyperon-nucleon ($YN$) system~\cite{Holz}. Specifically, this
means that one has used the same scheme (time-ordered perturbation
theory), the same type of processes, and vertex parameters (coupling
constants, cutoff masses of the vertex form factors) fixed already by
the study of those other reactions. 

\begin{figure}[t]
\vspace*{+1mm}
\centerline{\hspace*{3mm}
\psfig{file=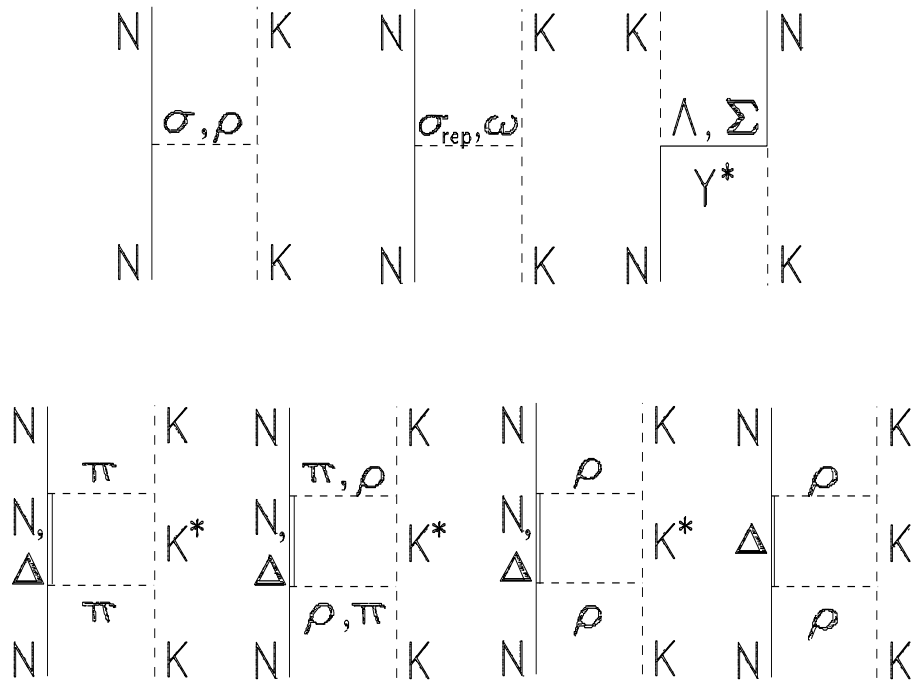,width=12.0cm,height=9.cm}}
\vspace*{+1mm}
\caption{Meson-exchange contributions to $KN$ scattering included
in the J\"ulich model I \cite{Juel2}. 
} 
\label{Diag}
\end{figure}

The diagrams considered for the $KN$ interaction are shown in 
Fig.~\ref{Diag}. Based on these diagrams 
the $KN$ potential $V$ is
derived, and the corresponding reaction amplitude
$T$ is then obtained by a solving a Lippmann-Schwinger type 
equation defined by time-ordered perturbation theory:
\begin{equation}
T = V + V G_0 T \ .
\label{LSE}
\end{equation}
From this amplitude phase shifts and observables can be obtained in 
the usual way.

As evident from Fig.~\ref{Diag},
the J\"ulich model contains not only single-meson
and baryon exchanges, but also higher-order box diagrams 
with $NK^*$, $\Delta K$ and $\Delta K^*$ intermediate 
states. Most vertex parameters involving the nucleon and the $\Delta$(1232)
isobar were taken over from the  Bonn $NN$ potential. 
The coupling constants at vertices involving strange baryons are fixed 
from the $YN$ model~\cite{Holz}. 
For the vertices involving mesons only, most coupling constants
have been fixed by SU(3) relating them to the empirical 
$\rho{\to}2\pi$ decay. Exceptions are the ${KK\sigma}$ and 
${KK\omega}$ coupling constants. The $\sigma$ meson, with a mass of 
about 600 MeV, is not considered as a genuine particle but as a simple
parametrization of correlated $2\pi$-exchange processes in the 
scalar-isoscalar channel. Therefore, its coupling strength cannot
be taken from symmetry relations. In the initial J\"ulich model 
\cite{Juel1,Juel3,Juel4} it was simply adjusted by a fit to the
$KN$ data. 
In a subsequent investigation \cite{Juel2} the $\sigma$(600) and also 
the elementary $\rho$ were replaced by a microscopic model for
correlated 2$\pi$ and $K\bar K$ exchange between kaon and nucleon,
in the corresponding scalar-isoscalar and vector-isovector
channels~\cite{Juel2}. Starting point for this was a model
for the reaction $N{\bar N}{\to}K{\bar K}$ with intermediate 2$\pi$
and $K\bar K$ states, based on a transition in terms of baryon, i.e.
$N$, $\Delta$, $\Lambda$ and $\Sigma$, exchange and a realistic
coupled channel $\pi\pi{\to}\pi\pi$, $\pi\pi{\to}K\bar K$ and
$K\bar K{\to}K\bar K$ amplitudes. The contribution in the $s$-channel 
is then obtained by performing a dispersion relation over the 
unitarity cut. 

Concerning the $\omega$-exchange it was found that a much 
larger strength than obtained from SU(3) would be required in order to
get sufficient short-range repulsion for a reasonable description of 
the $s$-wave $KN$ phase shifts~\cite{Juel1}. Thus, a phenomenological,
very short-ranged contribution was introduced in the potential 
(called $\sigma_{rep}$) which 
then allowed to achieve a satisfactory description of the $KN$ data 
\cite{Juel2}. Recently, it was shown that this phenomenological piece
can by substituted by contributions from genuine quark-gluon exchange
processes \cite{Hadjimichef}.

\begin{figure}[t]
\vspace*{-6mm}
\centerline{\hspace*{3mm}
\psfig{file=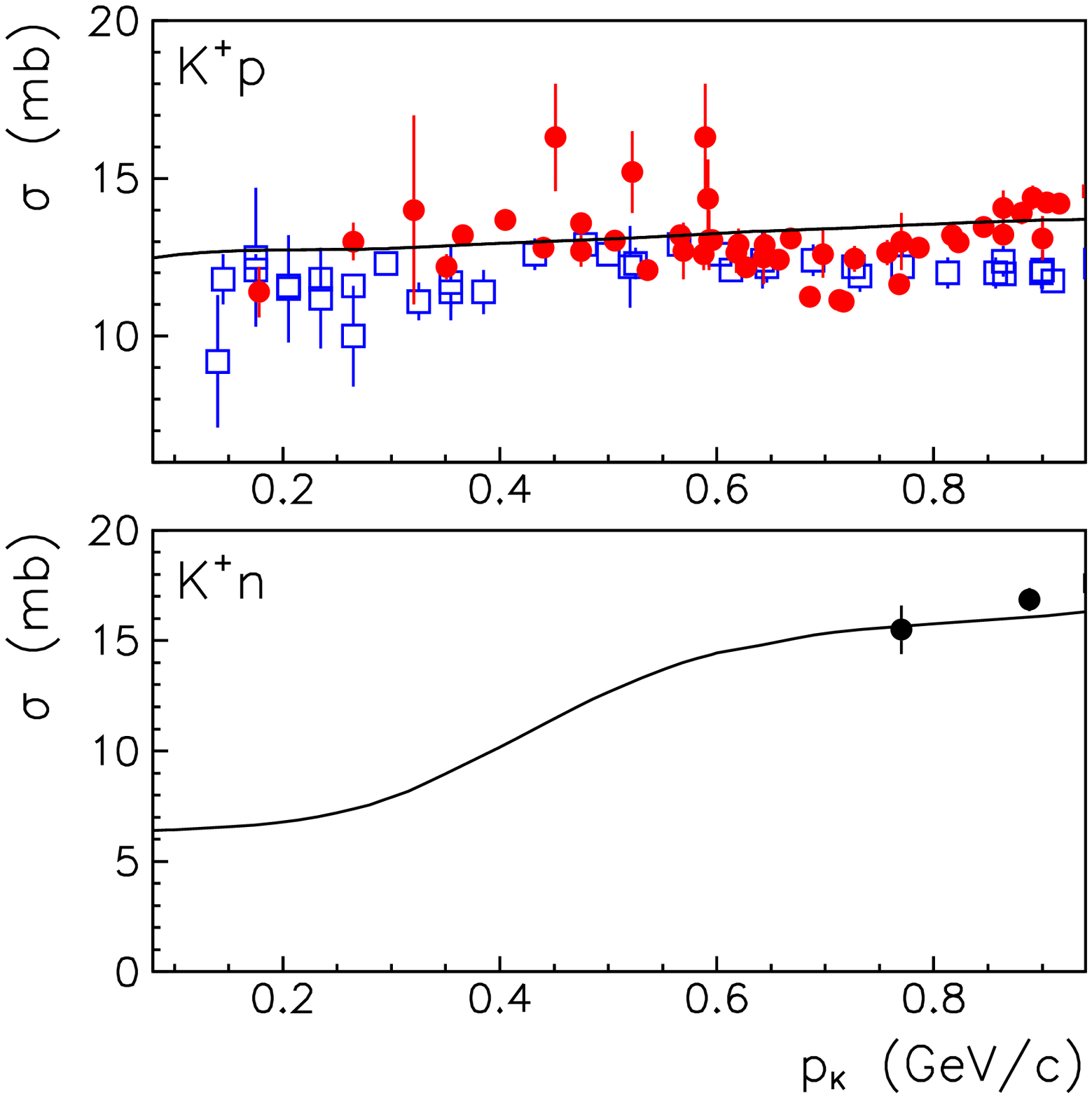,width=8.2cm,height=10.cm}\hspace*{-13mm}
\psfig{file=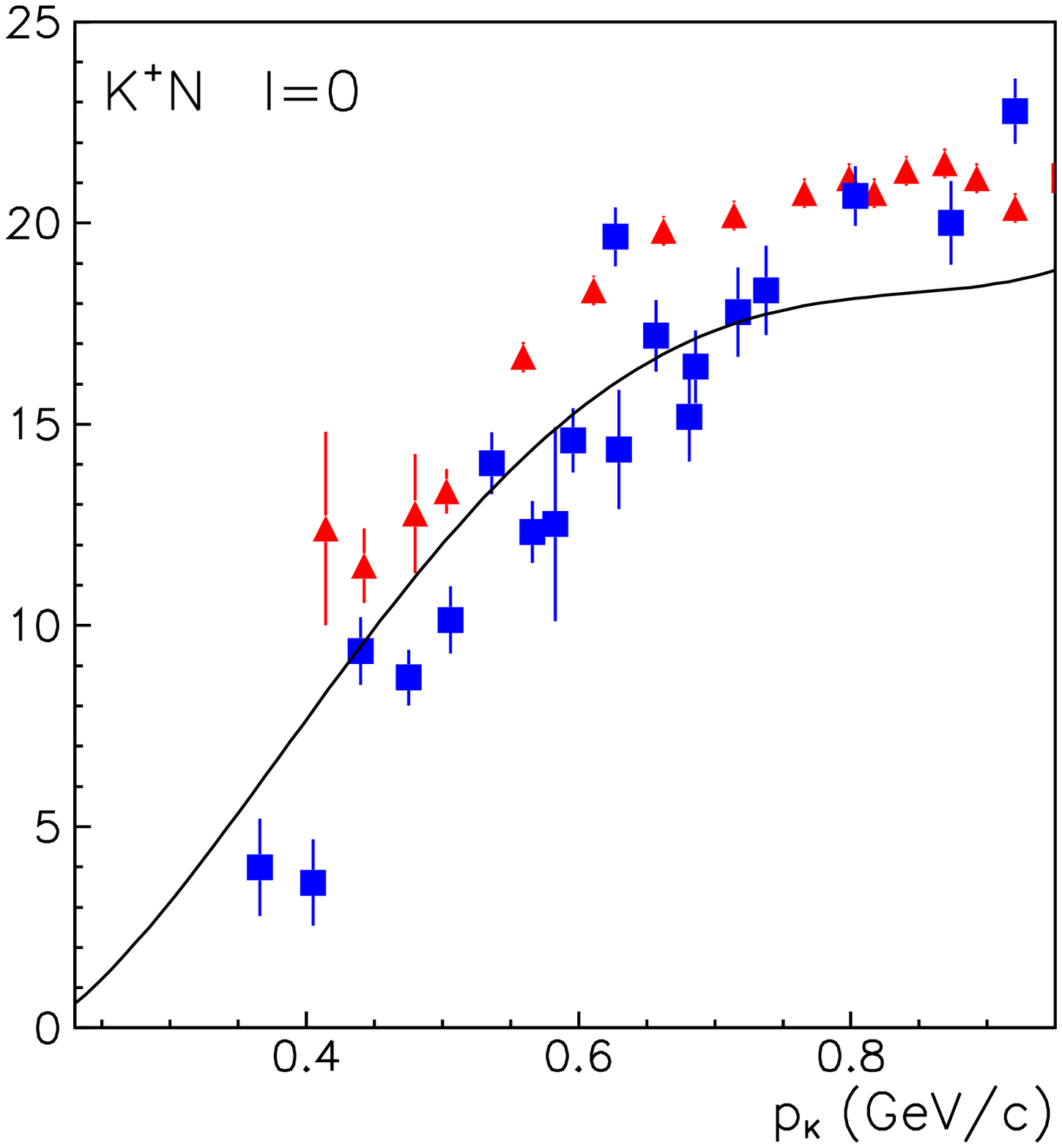,width=8.2cm,height=10.cm}}
\vspace*{-5mm}
\caption{The $K^+p$, $K^+n$ and $I{=}0$ cross sections as a function
of the kaon momentum. The circles are data for the total ($K^+p$, $K^+n$) 
cross section, while open squares are the elastic cross section, all
taken from Ref.~\cite{PDG}. 
The $I{=}0$ results are taken from Refs.~\cite{Carroll} (triangles)
and \cite{Bowen1,Bowen2} (squares). 
The solid lines are the results of the J\"ulich $KN$ model I \cite{Juel2}. 
}
\label{kade11}
\end{figure}

Since the results in Refs.~\cite{Juel2,Hadjimichef} indicate 
that the various $KN$ models presented by the J\"ulich group
yield a rather similar quantitative description of the $KN$ data 
we will employ here only one of those model. 
Specifically, we will use the $KN$ model I presented in
Ref.~\cite{Juel2}. Its parameters and the predictions for the
$KN$ phase shifts can be found in that reference, together
with a comprehensive comparision with empirical $KN$ scattering 
data, and therefore we do not reproduce them here. 
However, we want to illustrate some specific features of the
elementary $K^+N$ results which are relevant for the 
present investigation on the $K^+d$ reaction. 
Fig.~\ref{kade11} shows the $K^+p$ and $K^+n$
total and elastic cross sections as a function of the kaon momentum. 
The circles (squares) show the total (elastic) cross 
sections~\cite{PDG}. The solid lines are the results of the J\"ulich 
model for the total cross sections. 
The total and elastic cross sections for $K^+p$ scattering are 
practically identical below kaon momenta of $p_K{\simeq}$800 MeV/c, 
i.e. inelasticities which can occur from $p_K{\simeq}$520 MeV/c
onwards (due to the opening of the $KN\pi$ channel) are negligible,
and there is practically no energy dependence. 
The reaction channel corresponds to a pure $I{=}1$ state. It
is dominated by the $s$-wave amplitudes. Indeed, the available 
differential data for the $K^+p$ reaction indicate isotropic angular 
spectra for $p_K{<}$800 MeV/c, cf. Ref.~\cite{Juel2}. 

The elementary $K^+n$ data can only be obtained indirectly from a
study of the $K^+d$ reaction. The Particle Data Group~\cite{PDG}
lists only two data points for the total reaction cross section at low
momenta. The J\"ulich model indicates a pronounced momentum
dependence of the total $K^+n$ cross section. It is primarily
due to the $p$-wave contribution to the $I{=}0$ amplitude, in
particular the $P_{01}$ partial wave, which is substantial already for low kaon 
momenta \cite{Juel2}. Note that the $K^+n$ amplitude is 
given by (half of) the sum of the $I{=}0$ and $I{=}1$ isospin amplitudes. 
The total $I{=}0$ cross section is shown in the right
panel of Fig.~\ref{kade11} as a function of $p_K$. 
The J\"ulich model yields a reasonable reproduction of the data 
collected in Refs.~\cite{Carroll,Bowen1,Bowen2}. 
However, we want to emphasize that the 
$I{=}0$ $KN$ cross section was extracted from the $K^+d$ data 
at $p_K{<}800$ MeV/c by applying the
high energy Glauber formalism~\cite{Glauber,Franco}, which 
apparently introduces some ambiguities in the data evaluation. 
This point will be discussed in Sect.5. 

\section{Charge exchange and break-up reactions}

\subsection{The formalism}

A detailed and general description of the formalism for
two-body scattering of a spin-zero and a spin-1/2 particle is 
given by H\"ohler~\cite{Hoehler}. Thus, we provide here only a 
brief overview. 
The $KN$ reaction can be completely characterized by the 
spin-nonflip and spin-flip amplitudes $f$ and $g$ which, 
in terms of their partial wave projections, are given by
\begin{eqnarray}
f_I&=&\sum_{l=0}^\infty [(l+1)T^{l+}_I+lT^{l-}_I]\,\,
P_l(\cos\theta), \nonumber \\
g_I&=&\sum_{l=1}^\infty
\sin\theta\,\, [(T^{l+}_I-T^{l-}_I]\,\, P_l^\prime(\cos\theta) \ .
\label{part}
\end{eqnarray}
Here $l$ is the orbital angular momentum, and
$P_l$ and $P_l^\prime$ are Legendre polynomials and their
derivatives, respectively. Furthermore, $\theta$ is the scattering 
angle in the $KN{\to}KN$ center of mass system and $T_I^{l\pm}$ is 
the PW amplitude for the total angular momentum $l{\pm}1/2$ and 
isospin $I$, where the latter can be 0 or 1 for the
$KN$ system. The amplitudes of the three possible physical 
$KN$ reactions are related to those in Eq.~(\ref{part}) by
\begin{eqnarray}
K^+p&{\to}&K^+p: \ \  F_p{=}f_1+f_C, \,\, G_p{=}g_1, 
\label{kaplup} \\
K^+n&{\to}&K^+n: \ \  F_n{=}1/2(f_1+f_0), \,\, G_n{=}1/2(g_1+g_0), \\
K^+n&{\to}&K^0p: \ \  F_{ex}{=}1/2(f_1-f_0), \,\,
G_{ex}{=}1/2(g_1-g_0),
\label{eq5}
\end{eqnarray} 
where $f_C$ stands for the Coulomb amplitude. The scattering
amplitude for each of those reactions is then given in terms of the
corresponding quantities $F$ and $G$ by 
\begin{eqnarray}
{A}=F+i{\mbox{\boldmath $\sigma$}}\cdot [{\bf k_0}\times{\bf k}]\,G,
\label{aplto}
\end{eqnarray}
where ${\bf k_0}$, ${\bf k}$ defines the direction of initial and final
kaon, respectively and {\boldmath $\sigma$} is the Pauli matrix.
The two-body scattering cross section and the polarization are given by
\begin{eqnarray}
\frac{d\sigma}{d\Omega}{=}|A|^2{=}|F|^2{+}|G|^2, \ \ 
P{=}\frac{2\, {\rm Im}(FG^\ast)}{|F|^2{+}|G|^2} \ .
\end{eqnarray}
It is clear that the amplitudes $f_1$ and $g_1$ can be obtained from an 
analysis of the $K^+p{\to}K^+p$ elastic scattering data, while the
determination of $f_0$ and $g_0$ relies on experiments with a deuteron
target. 

The differential cross section for the reaction $K^+d{\to}K^+np$ 
is given in the impulse approximation by 
\begin{eqnarray}
\frac{d\sigma}{d\Omega} &=& \int |A_d(k_0,k,q)|^2\,\,
\delta^4(k_0{+}P{-}k{-}p{-}q)\,\,
\frac{k^2dk}{E_K}
\frac{d^3p}{E_p}\frac{d^3q}{E_q}, \nonumber \\
A_d &=& A_n(k_0,k,q)\,\, \Psi({\bf p}) + A_p(k_0,k,p)\,\, \Psi({\bf q}) \ , 
\label{form7}
\end{eqnarray}
where $k_0$, $k$, $P$, $q$, and $p$ are the momenta of the initial and
final kaon, of the deuteron and of the final neutron and proton,
while $E_K$, $E_p$ and $E_q$ are the total energies of the particles in
the final state. Furthermore, $\Psi$ is the deuteron wave function in
momentum space: 
\begin{eqnarray}
\Psi({\bf q}) = \psi_0 (q) + \frac{1}{\sqrt{2}} 
\left( 3 \frac{({\bf S} \cdot {\bf q})^2}{q^2} - 2 \right) 
\psi_2 (q) \ , \ \   
{\bf S} = \frac{1}{2} ({\bf \sigma_1} + {\bf \sigma_2}) \ .
\end{eqnarray}
The scattering amplitude is taken at 
$t{=}(k{-}k_0)^2$ four-momentum transfer squared and at the squared
invariant energy $s{=}(k{+}q)^2$ or $s{=}(k{+}p)^2$, respectively.
We should mention that in this expression not only effects from 
kaon rescattering and the ($NN$) final state interaction are 
neglected, in line with the impulse approximation, 
but also corrections due to the deuteron binding.

Note that, in principle, the 
$K^+n$ and $K^+p$ amplitudes enter off-shell in Eq.~(\ref{form7}) 
because the interacting nucleon is off its mass shell. 
However, in an integration over the three-body phase space
the dominant contribution to the cross section comes from the
kinematics near zero spectator-momentum, where $A_n$ and $A_p$, 
respectively, are close to their on-shell values and, therefore,
in general one uses only the on-shell amplitudes. 
 
In the application of the impulse approximation to the reaction
$K^+d{\to}K^+np$ and  $K^+d{\to}K^0pp$
one usually makes additional simplifications. 
First one neglects the $d$-wave component of the deuteron wave
function. 
Secondly, one assumes that the energy dependence of the scattering 
amplitude is smooth within the range of 
integration of Eq.~(\ref{form7}).
Finally, one assumes that the elementary two-body $KN$ amplitude
enters into the three-body reaction in a kinematics where 
the scattered nucleon (and accordingly also the spectator)
has zero momentum in the laboratory system. The latter two
assumption allow to factorize $|A_d|^2$ out of
the integral of Eq.~(\ref{form7}). 
Implementing these assumptions Stenger et al. \cite{Stenger} derived 
the following relations for the reactions $K^+d{\to}K^0pp$
and $K^+d{\to}K^+np$ 
\begin{eqnarray}
\frac{d\sigma}{d\Omega}(K^+d{\to}K^0pp){=}\biggr(|F_{ex}|^2{+}\frac{2}{3}
|G_{ex}|^2\biggr)I_{ppt}(\theta)+\frac{1}{3}|G_{ex}|^2I_{pps}, \label{xex}\\
\frac{d\sigma}{d\Omega}(K^+d{\to}K^+np){=}
\biggr(|F_p|^2{+}|F_n|^2{+}\frac{2}{3}|G_p|^2{+}\frac{2}{3}|G_n|^2
\biggr)I_{npt} \nonumber \\
+\frac{1}{3}\biggr(|G_p|^2{+}|G_n|^2\bigg)I_{nps}
{+}2{\rm Re}\biggr(F_n^\ast
F_p+\frac{2}{3}G_n^\ast G_p\biggr)J_{npt} \nonumber \\
{-}\frac{2}{3}{\rm Re}\biggr(G^\ast_nG_p\biggr)J_{nps}, 
\label{xsect}
\end{eqnarray}
where $I$ and $J$ are the so-called deuteron inelastic form factors. Their
subscripts specify whether the final $pp$ or $np$ pair is in a singlet
($s$) or triplet ($t$) state, respectively. In the plane wave
approximation the six ($I$ and $J$) form factors reduce to 
\begin{eqnarray}
I_{pps}{=}I_0{+}J_0, \,\,\, I_{ppt}{=}I_0{-}J_0, \,\,\,
I_{nps}{=}I_0,\,\,\, I_{npt}{=}I_0, \,\,\, 
J_{nps}{=}J_0, \,\,\, J_{npt}{=}J_0,
\label{nnstate}
\end{eqnarray}
where 
\begin{eqnarray}
I_0{=}D\int \frac{u^2(p){+}u^2(q)}{2}\,\,
\delta^4(k_0{+}P{-}k{-}p{-}q)\,\,
\frac{k^2dk}{E_K}
\frac{d^3p}{E_p}\frac{d^3q}{E_q}, \\
J_0{=}D\int u(p)u(q)\,\,
\delta^4(k_0{+}P{-}k{-}p{-}q)\,\,
\frac{k^2dk}{E_K}
\frac{d^3p}{E_p}\frac{d^3q}{E_q} \ .  
\end{eqnarray}
Here the kinematical factor $D$ accounts for the transformation of the kaon
scattering angle from laboratory $K^+d$ system to the center of mass
frame of the $K^+N$ two-body system. The scattering amplitudes $F$ and 
$G$ are evaluated for the stationary spectator nucleon, i.e. 
\begin{eqnarray}
D{=}\frac{E_K}{k^2}\,\, \left .\frac{d(E_K+E_q)}{dk}\right|_{p{=}0} \ .
\label{trans}
\end{eqnarray}

Let us emphasize here that Eq.~(\ref{nnstate}) is valid only in the
impulse approximation, which corresponds to a plane wave 
approximation for the wave function of the two final nucleons. 
For interacting nucleons the $NN$ wave function is different for 
the singlet and triplet states. 
Moreover, the $NN$ interaction in 
the final state might effect the forward scattering angles, specifially 
because this singlet interaction is rather strong~\cite{Stenger}.  

In our calculations we use the $KN$ amplitude of the J\"ulich model I
\cite{Juel2} and the deuteron wave function from the recent 
charge-dependent Bonn $NN$ potential~\cite{Machleidt}. We integrate the
two-body amplitude over the full three-body phase space, i.e. we do
not factorize the reaction amplitude. 
The calculations are
performed in the deuteron rest frame. That allows us to compare 
our results directly to the majority of the experimental data which 
are naturally given in the laboratory system. 
In the few other cases we compare our results in the
$K^+N$ center-of-mass frame by performing a transformation into the
two-body system under the assumption that the spectator nucleon is at rest,
i.e. following Eq.~(\ref{trans}).
Since the published data are given in different frames 
we compile the relevant kinematical relations in an Appendix
for the convenience of the reader.

\subsection{The reaction { ${\rm K^+d{\to}K^+np}$}}

There are several experiments for the reaction $K^+d{\to}K^+np$
and data are available for kaon momenta from 342 MeV/c onwards. 
The basic difficulty in comparing these experiments with our 
$KN$ model comes from the reaction kinematics, which is not 
explicitely described in most of the papers. 
In general the available data are differential reaction cross 
sections as a function of the $K$-meson scattering angle, 
which are given either in the laboratory (deuteron rest) system or in 
the quasi two-body $KN$ center-of-mass frame. 
The momentum of the spectator and/or of the scattered nucleon are
often not specified or not measured explicitely.  Thus, the
reaction can involve kaon scattering on both the proton and neutron and 
therefore we add the corresponding amplitudes according to 
Eq.~(\ref{form7}). 
This amplitude is then integrated over the full three-body
phase space and without any cuts on the momenta of the final particles.
\begin{figure}[t]
\vspace*{-6mm}
\centerline{\hspace*{3mm}\psfig{file=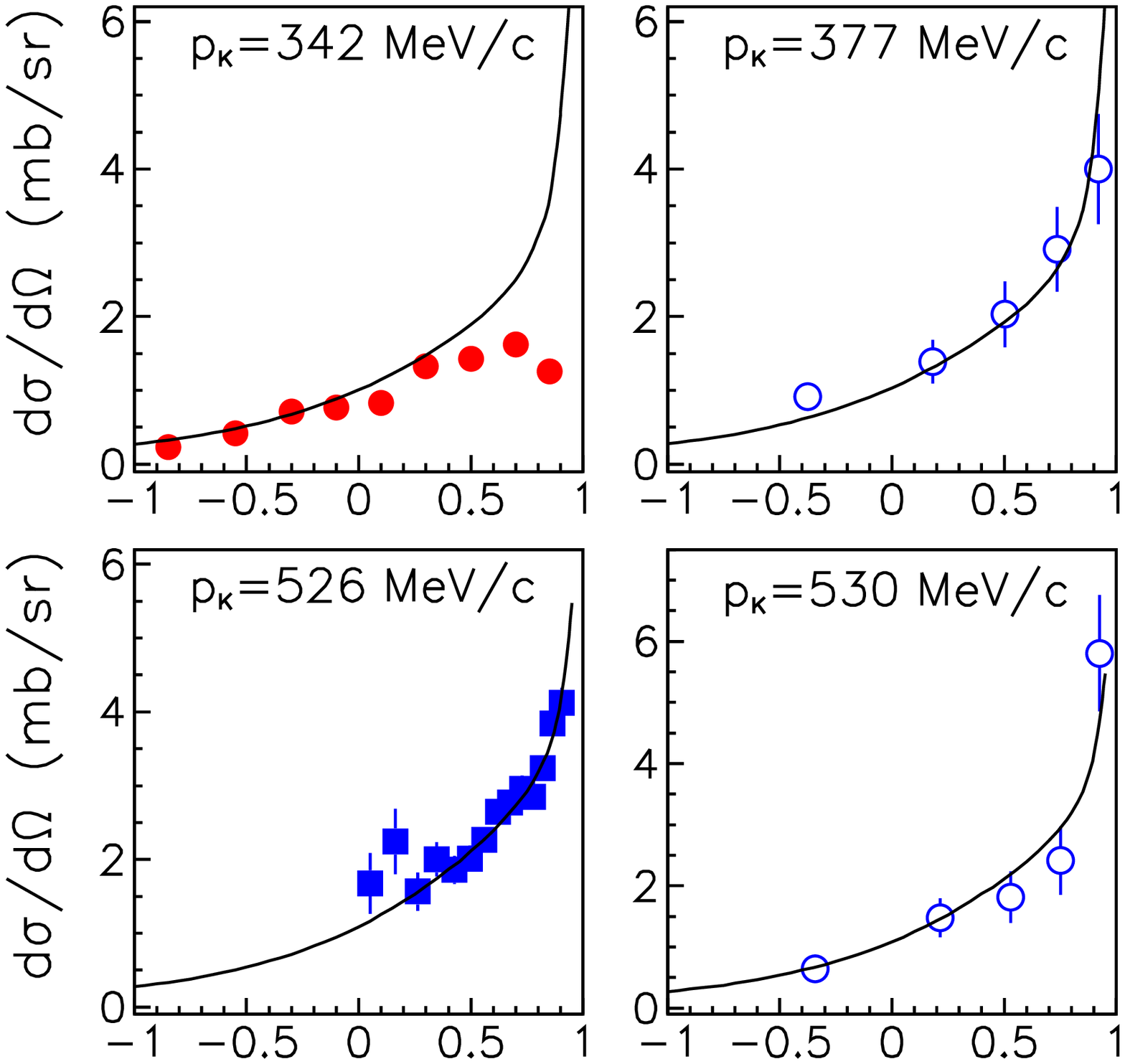,width=8.3cm,height=10.7cm}
\hspace*{-15mm}\psfig{file=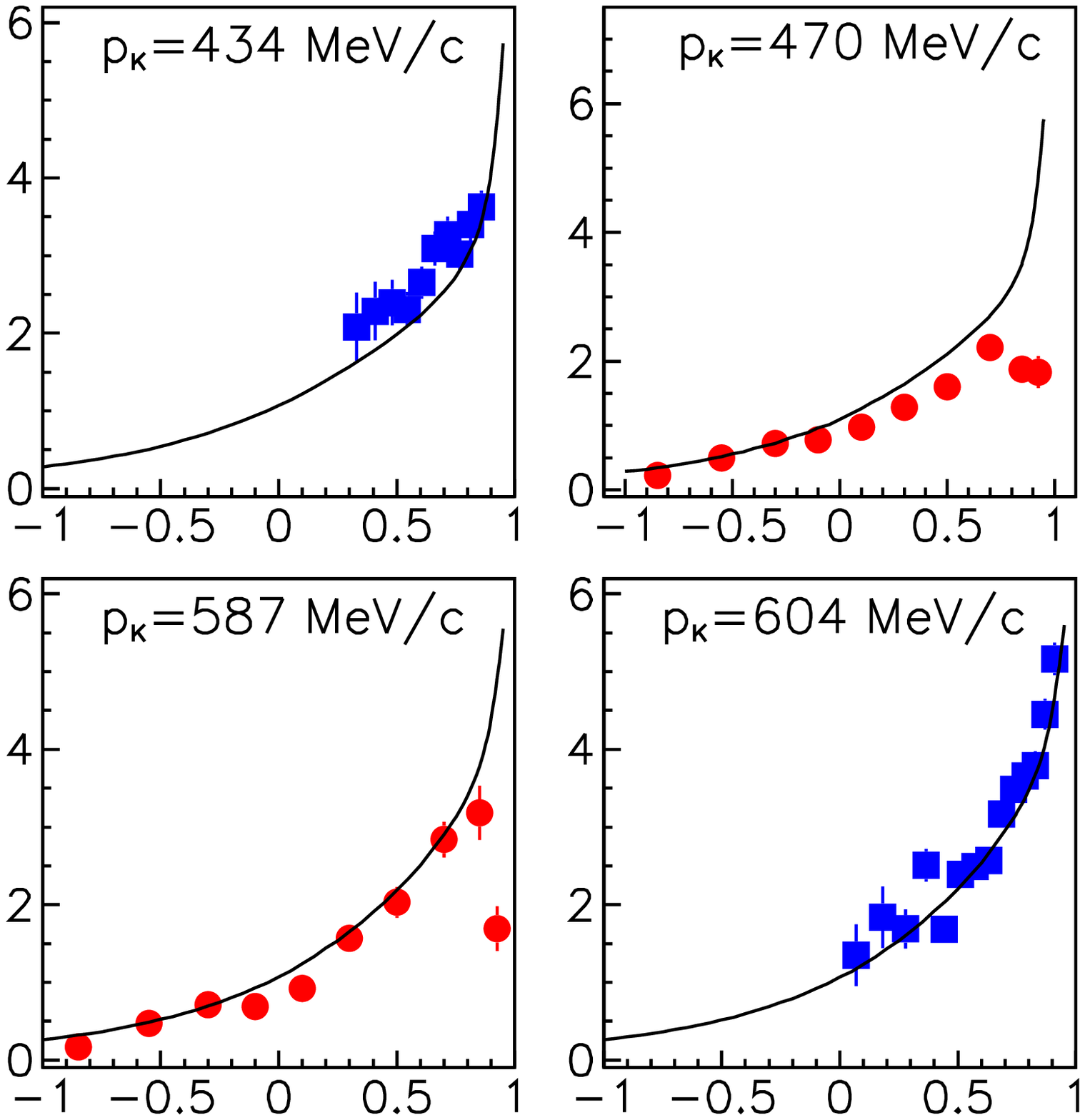,width=8.3cm,height=10.7cm}}
\vspace*{-17mm}
\centerline{\hspace*{3mm}\psfig{file=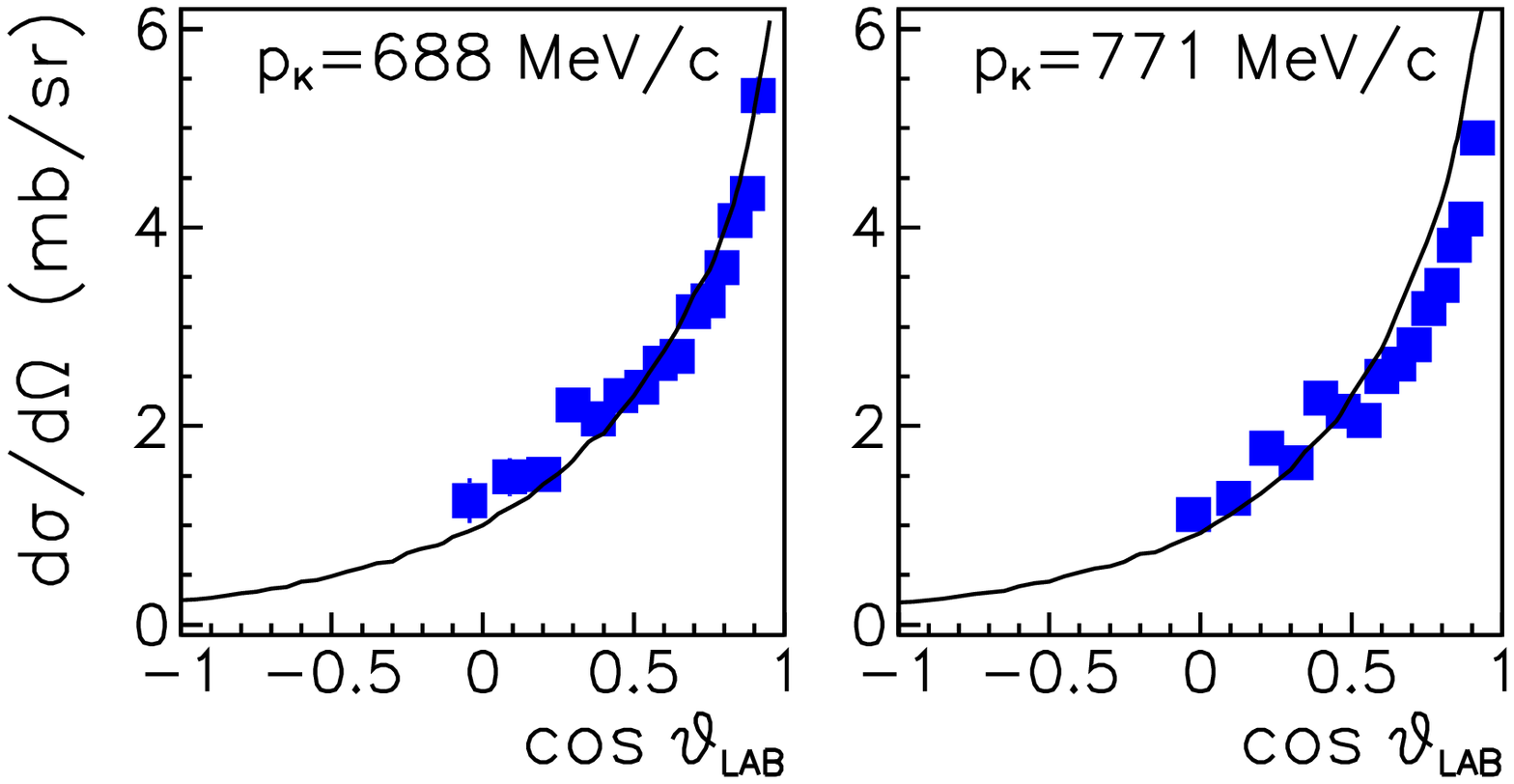,width=8.3cm,height=10.7cm}}
\vspace*{-50mm}
\caption{The $K^+d{\to}K^+np$ differential cross section for different
incident kaon momenta as a function of the kaon scattering angle in the
laboratory system. 
The data are from Refs.~\cite{Stenger} (open circles), 
\cite{Glasser} (filled circles), and \cite{Damerell} (squares). 
The solid lines show the result of our model calculation.} 
\label{kade1}
\end{figure}
This might be inadequate since the
momenta and angles of the final proton and neutron can be limited due
to the detector acceptance. However, without corresponding 
information such kinematical constraints cannot be implemented in
the three-body integration.

Results of our calculations are presented in Fig.~\ref{kade1}, together 
with data from Refs.~\cite{Stenger,Glasser,Damerell}. 
The differential cross
sections are shown as a function of the kaon scattering angle in the
laboratory system. We use this frame in order to avoid ambiguities in the
transformation from the deuteron rest frame, where the experiment was done,
to the effective $KN$ center-of-mass system. Note that this ambiguity
is discussed in different experimental papers in detail and there
are various prescriptions for that transformation, for instance the
one given by Eq.~(\ref{trans}).

The result at the lowest energy is due to Glasser et al.~\cite{Glasser}. 
In this experiment the differential $K^+d{\to}K^+np$ cross section 
was measured by identification of the 2-prong
events. To distinguish between the $K^+d{\to}K^0pp$ reaction, which
also has two charged particles in the final state, an additional
criteria on not-associated $V$ tracks from the $K^0{\to}\pi^+\pi^-$
decay was imposed. 
Some misidentification of the reactions $K^+d{\to}K^+np$ and
$K^+d{\to}K^+d$ comes from the proton and final deuteron
tracks saturation and difficulties to distinguish between them. 
While our calculation yields a good overall description of the data 
-- at 342 MeV/c as well as at 587 MeV/c -- 
there is obviously a discrepancy at forward angles to which we
will come back below. 

The reaction $K^+d{\to}K^+np$ was also measured by Stenger et 
al.~\cite{Stenger}. However, in their experiment it was not possible to
separate the contributions from elastic $K^+d{\to}K^+d$ scattering and
from $K^+d{\to}K^+np$ break-up. Therefore, the data shown by squares in
Fig.~\ref{kade1} are the sum of the coherent and $K^+d{\to}K^+np$
cross sections. The solid lines in Fig.~\ref{kade1} are model predictions
for the genuine reaction $K^+d{\to}K^+np$. Obviously those results are 
already in pretty good agreement with the data reported
in Ref.~\cite{Stenger}. Thus, there is not much room left at
forward angles for possible contributions from $K^+d{\to}K^+d$ scattering. 
But, as we will show later, the coherent cross section is indeed very small 
and it strongly depends on the scattering angle due to the deuteron form 
factor. 

Another $K^+d{\to}K^+np$ experiment was performed by Damerell et
al.~\cite{Damerell}. The reaction was identified by detecting the
$K^+$-meson using a time-of-flight (TOF) spectrometer. 
The squares in Fig.~\ref{kade1} show data
that represent also the sum of the $K^+d{\to}K^+d$ and $K^+d{\to}K^+np$
differential cross sections. Again in this case our calculation
for the $K^+d{\to}K^+np$ reaction alone is in good agreement 
with the experimental results.

Let us come back to the data by Glasser et al. where 
the experiment shows a forward suppression, contrary to the theory. 
In this context let us recall that while comparing our $K^+d{\to}K^0pp$
calculations~\cite{Sibirtsev1} with the charge exchange data we did
not detect any discrepancy at forward angles. 
On the other hand, from Eq.~(\ref{xex}) one can see that the singlet $pp$
form factor $I_{pps}$ enters only via the spin-flip term $G_{ex}$, 
which itself is small and vanishes at forward angles. 
The spin non-flip part of the
amplitude, $F_{ex}$, is multiplied by the form factor $I_0{-}J_0$ of
Eq.~(\ref{nnstate}) that originates from
the spin triplet $pp$ interaction which also vanishes at forward
direction. The $K^+d{\to}K^0pp$ charge exchange differential 
cross section is thus suppressed at forward angles, which is clearly
seen in our results in Ref.~\cite{Sibirtsev1} that are also discussed 
in the next subsection. 
Therefore, it would be difficult to see any inadequacy of the 
employed elementary $KN$ amplitudes or of the impulse approximation
at small angles because the corresponding model results are
automatically reduced.

For the $K^+d{\to}K^+np$ reaction the situation is somewhat
different. Since the spin
non-flip amplitude $F$ dominates the reaction and both ($I_0$ and $J_0$)
form factors do not vanish at forward angles, there is no
suppression of the differential cross section at forward angles 
in the model calculation. Therefore, the discrepancy with the  
Glasser data \cite{Glasser} in forward direction, cf. Fig.~\ref{kade1}, 
could indeed be a signal for a failure of the impulse 
approximation, say. On the other
hand, and may be more likely, it could be simply due to ambiguities 
in the data evaluation, specifically, because we do get nice
agreement with the forward data within the same energy range 
provided by other groups \cite{Stenger,Damerell}. 
In this context we would like to mention that the authors of 
Ref.~\cite{Glasser} did not include their $K^+d{\to}K^+np$ 
data in their own $KN$ partial wave analysis nor did they 
confront the results of that analysis with the measured 
$K^+d{\to}K^+np$ differential cross sections. 


\begin{figure}[t]
\vspace*{-6mm}
\centerline{\hspace*{3mm}\psfig{file=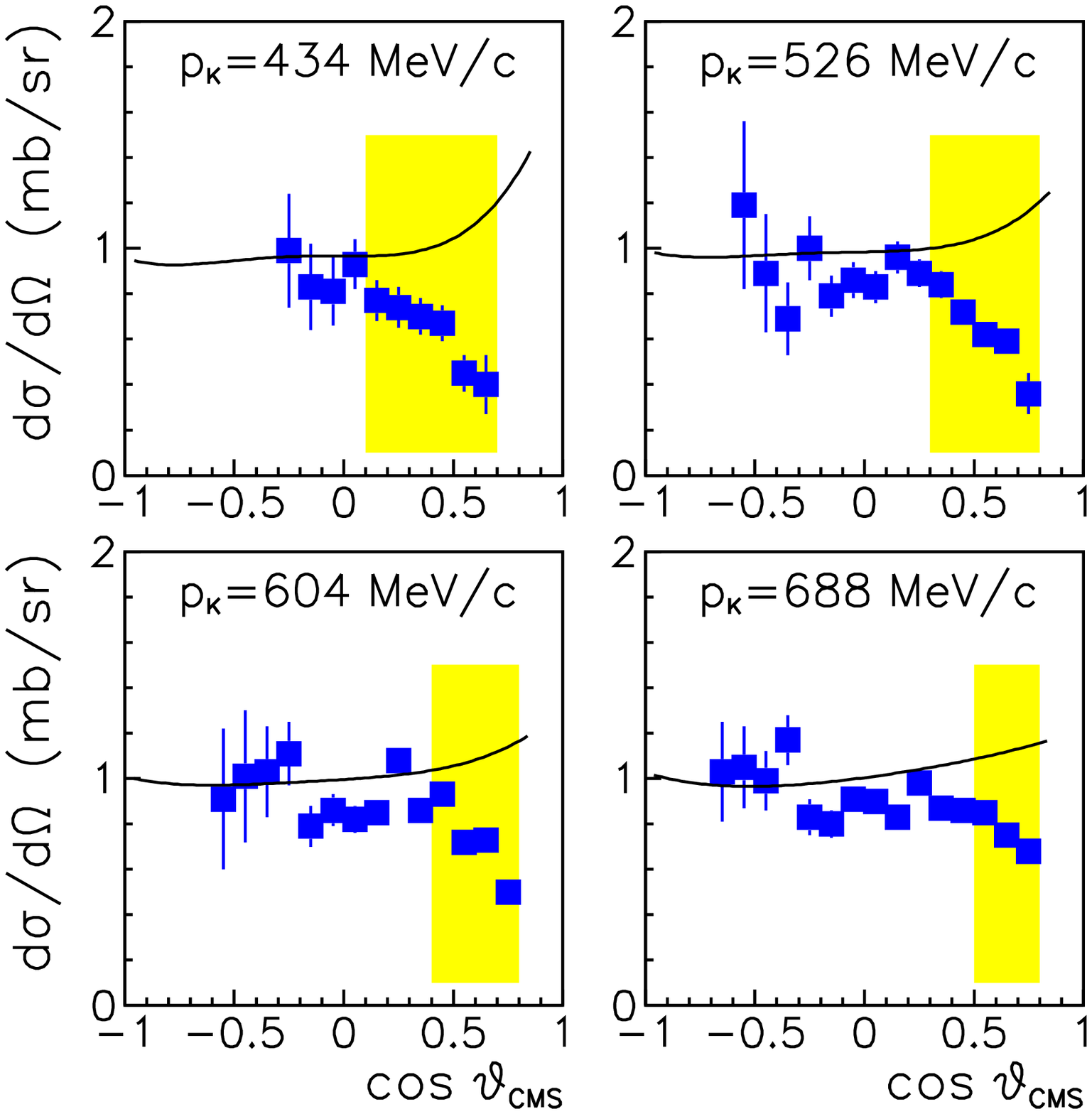,width=8.3cm,height=10.7cm}\hspace*{-15mm}\psfig{file=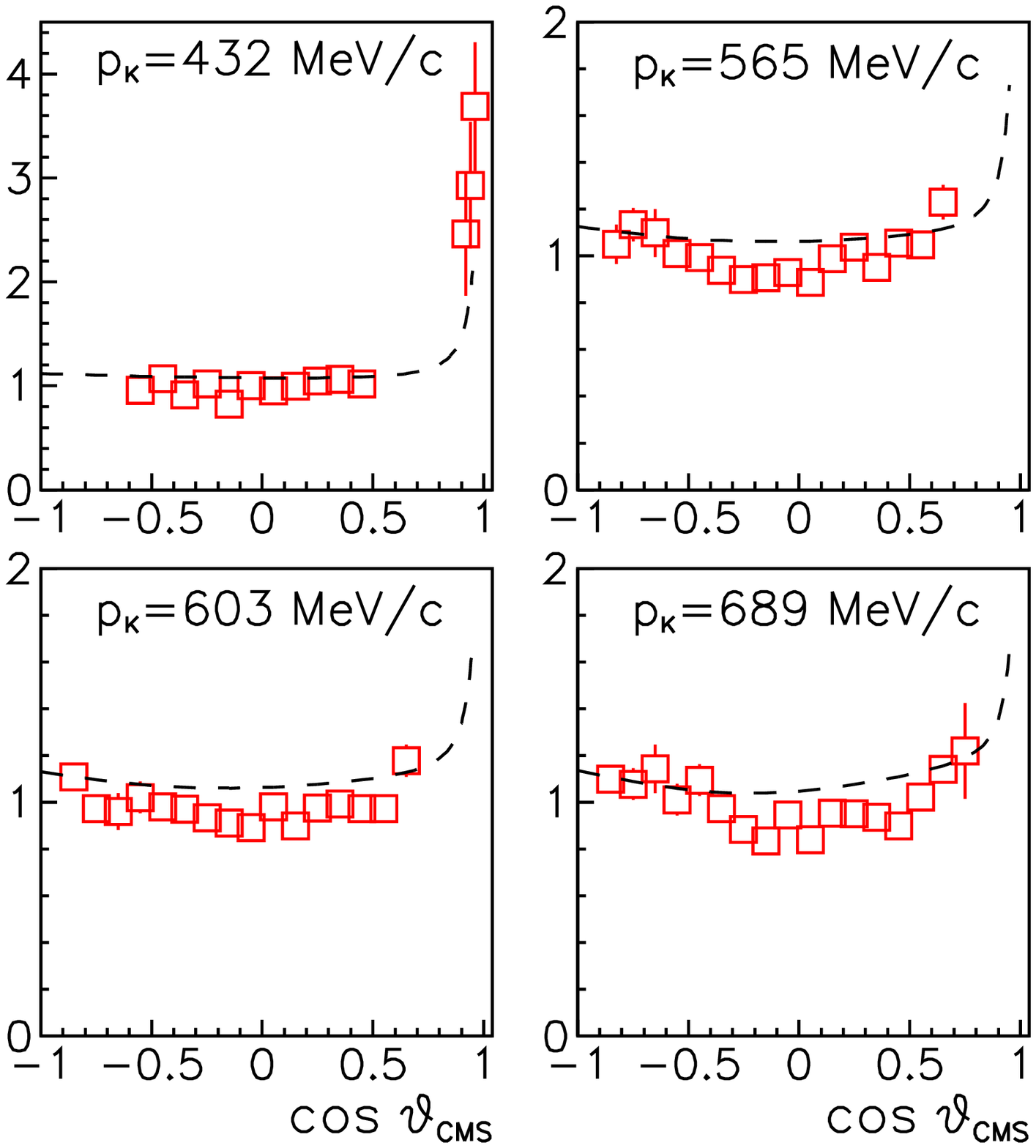,width=8.3cm,height=10.7cm}}
\vspace*{-5mm}
\caption{The $K^+p{\to}K^+p$  differential cross section for different
incident kaon momenta as a function of the kaon scattering angle in the
center of mass system. The filled squares show the results extracted from 
the $K^+d{\to}K^+pn$ reaction~\cite{Damerell} with reconstructed neutron
spectator. The shaded areas indicate the angular range where data 
analysis was considered to be ambiguous by the authors.
The open squares are the differential cross sections measured
on a hydrogen target~\cite{Adams}. The solid lines show our 
calculation for the reaction $K^+d{\to}K^+pn$ (assuming that
the $n$ is the spectator), while the dashed lines indicate our 
(elementary) $K^+p{\to}K^+p$ results.} 
\label{kade3}
\end{figure}

Besides data on the reaction $K^+d{\to}K^+np$ one can also find
experimental results for the elementary $K^+n \to K^+n$ process
in the literature~\cite{Damerell,Giacomelli2}. In order to obtain
such data one has to invoke the spectator model and one has to
isolate those $K^+d{\to}K^+np$ events with either a proton or a 
neutron as spectator, i.e. where either the proton or the 
neutron momentum fulfils the spectator condition 
$p_p < p_n$ or $p_n < p_p$, respectively.  
However, in case of the data of Damerell et al.~\cite{Damerell} it
remains to some extent unclear how the experimental separation
between the reactions with a spectator proton or neutron was done
since, as already said earlier, the direction and momentum of only one
of the outgoing particles was measured and it was not possible to 
specify exactly the kinematics of each event. 
Nonetheless their analyis had the ambitious goal to extract also 
$K^+p$ cross sections from $K^+d{\to}K^+pn$ events with a spectator 
neutron and compare them with the free differential cross section 
for $K^+p$ scattering. Such a comparison constitutes a crucial test 
for extracting (elementary) $KN$ cross sections from measurements 
on a deuteron target, provided that it is performed with a 
sensible data set. 

\begin{figure}[t]
\vspace*{-6mm}
\centerline{\hspace*{3mm}\psfig{file=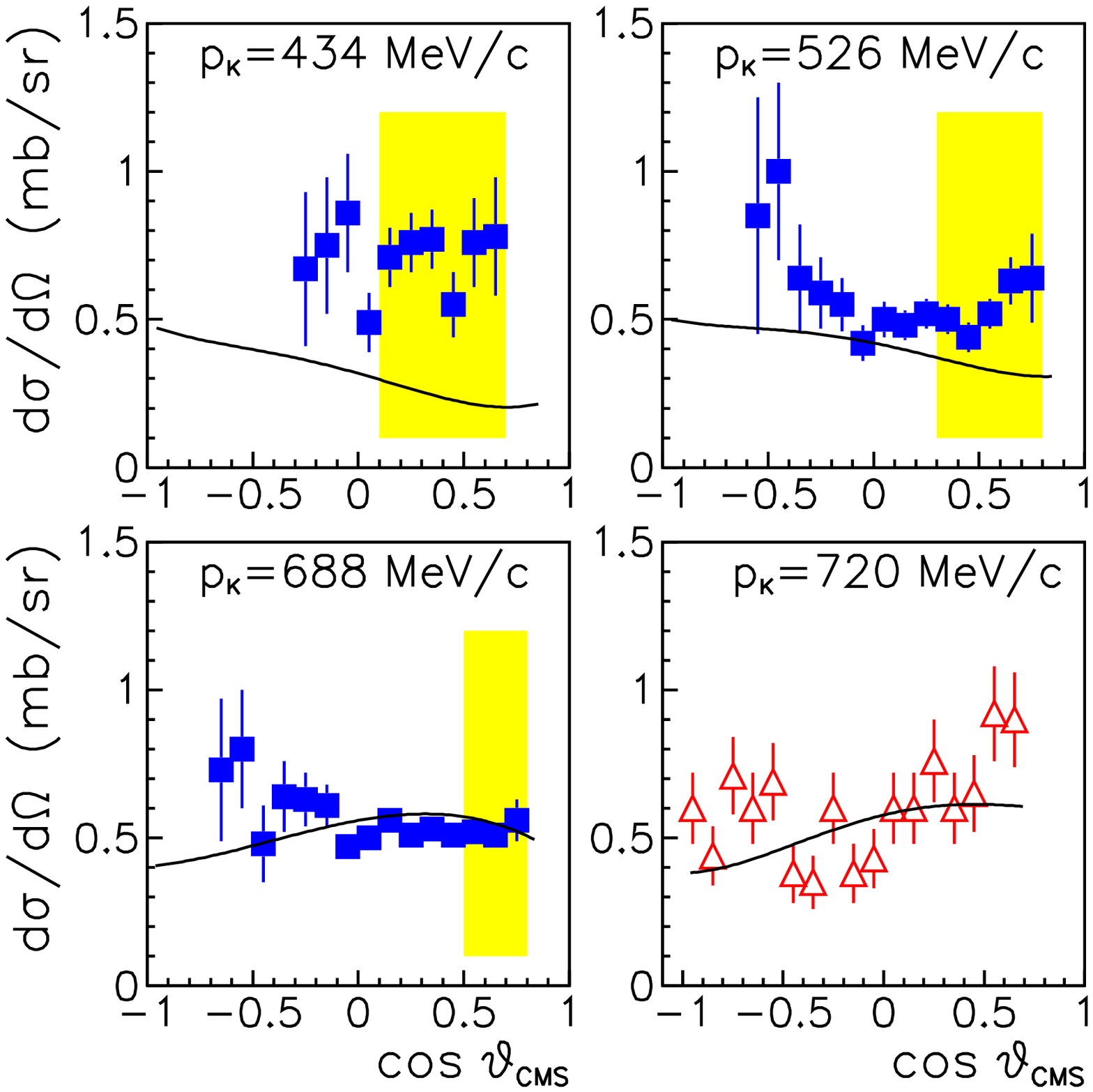,width=8.2cm,height=10.7cm}\hspace*{-13mm}\psfig{file=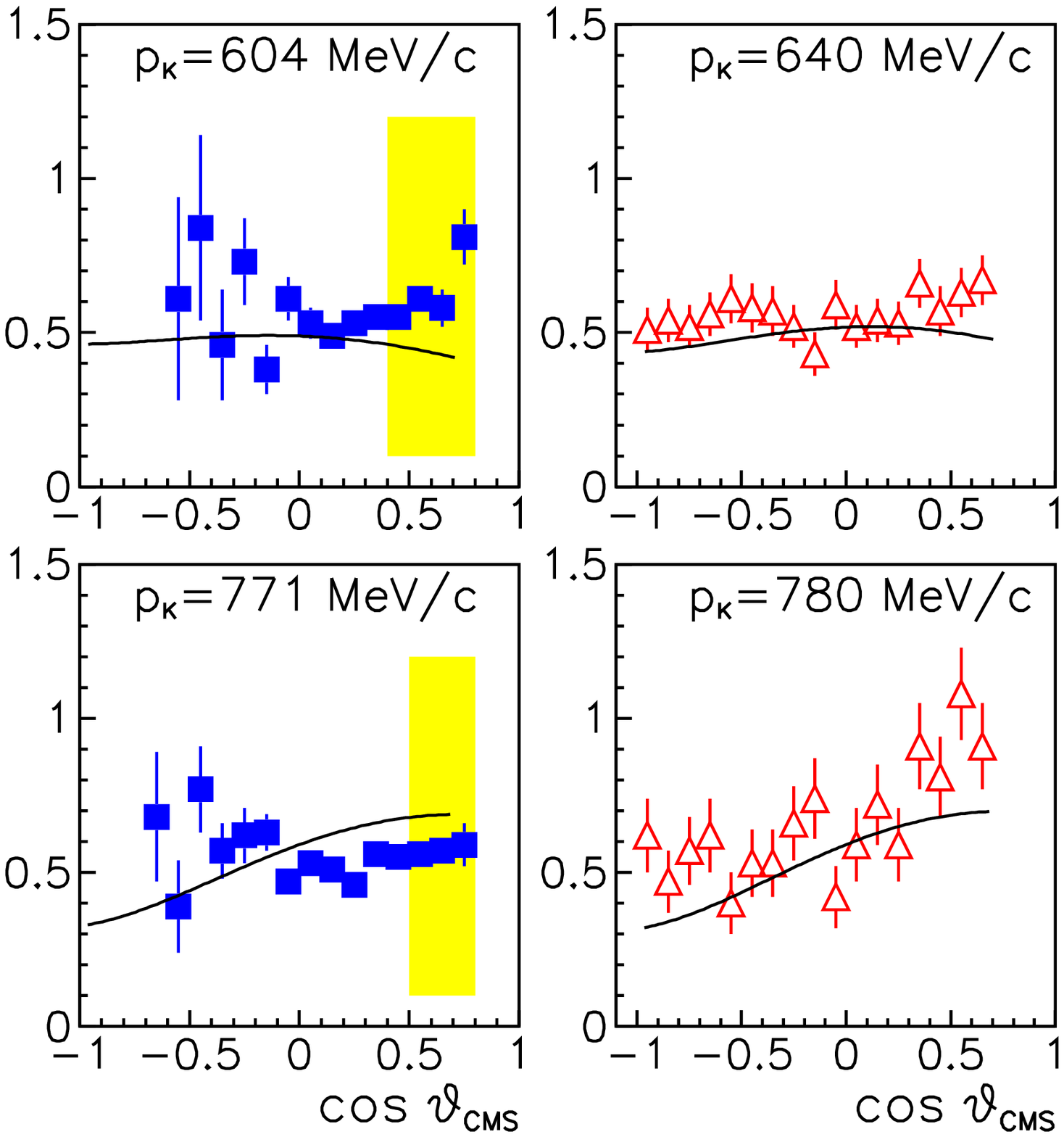,width=8.2cm,height=10.7cm}}
\vspace*{-5mm}
\caption{The $K^+n{\to}K^+n$  differential cross section for different
incident kaon momenta as a function of the kaon scattering angle in the
center of mass system. The experimental results were extracted from the
$K^+d{\to}K^+np$ reaction with reconstructed neutron
spectator; squares show the data from Ref.~\cite{Damerell},
while the triangles were taken from Ref.~\cite{Giacomelli2}. The shaded areas 
indicate the angular range where the data analysis was considered to be ambiguous
by the authors.
The solid lines show our calculations for the $K^+d{\to}K^+np$ reaction
(assuming that the $p$ is the spectator). 
} 
\label{kade9}
\end{figure}

The $K^+p{\to}K^+p$ differential cross sections 
extracted by Damerell et al. from their $K^+d{\to}K^+np$
data are shown by the filled squares in
Fig.~\ref{kade3}. The shaded areas indicate the region where, 
according to the authors~\cite{Damerell}, possible uncertainties 
in the treatment of the final proton and effects due to coherent 
scattering affect the data analysis.
The solid lines in Fig.~\ref{kade3} show our calculations for the
$K^+d{\to}K^+np$ reaction with a spectator neutron, i.e. with the 
amplitude $A_p$ of Eq.~(\ref{aplto}). The calculations and data are shown 
in the $K^+N$ c.m. frame in order to enable a comparision with 
data as well as calculations for the free $K^+p$ reaction,
which are presented at the right side of Fig.~\ref{kade3}.
Those data (open squares in Fig.~\ref{kade3}) are
$K^+p{\to}K^+p$ differential cross sections measured~\cite{Adams} with
a hydrogen target and at kaon momenta close to that studied in the  
$K^+d{\to}K^+np$ experiment~\cite{Damerell}. The corresponding model
predictions are given by the dashed line. 
Obviously both sets of data are in rough agreement outside of the shaded 
areas. But there is indeed a strong disagreement at forward angles. 
 
\begin{figure}[t]
\vspace*{-6mm}
\centerline{\hspace*{3mm}\psfig{file=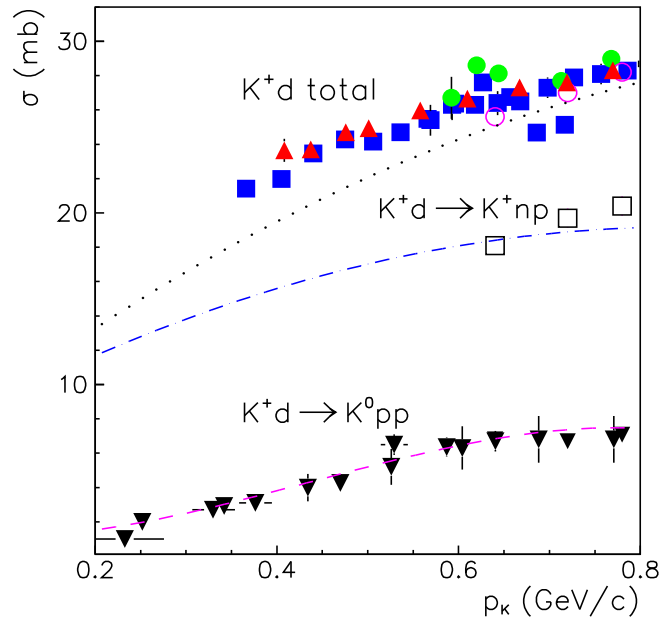,width=9.6cm,height=9.4cm}}
\vspace*{-5mm}
\caption{The total $K^+d$ cross section and the integrated $K^+d{\to}K^+np$ 
and $K^+d{\to}K^0pp$ reaction cross sections as a function of the kaon momentum. 
The dashed and dash-dotted lines are the results of our model 
calculations for the $K^+d{\to}K^0pp$ and $K^+d{\to}K^+np$ cross sections,
respectively, while the dotted line is their sum.
Data for the $K^+d$ total cross section are from Refs.~\cite{Bowen1,Bowen2}
(filled squares), \cite{Carroll} (filled triangles),
\cite{Bugg1} (filled circles) and \cite{Giacomelli3} (open circles). 
The
$K^+d{\to}K^+np$ data are from \cite{Giacomelli3} (open squares), 
while the $K^+d{\to}K^0pp$ cross section are from \cite{Damerell,Slater}
(inverse triangles). 
} 
\label{kade7}
\end{figure}

Fig.~\ref{kade9} contains the $K^+n{\to}K^+n$ differential cross section
extracted from the $K^+d{\to}K^+np$ for different kaon momenta and as
a function of the $K^+$-meson scattering angle in the $KN$ c.m. system. 
The squares are the results from Ref.~\cite{Damerell},
while triangles show the measurements from Ref.~\cite{Giacomelli2}. 
Note that in the latter experiment the spectator condition for the proton
was imposed in selecting the events. Specifically, only those events 
were accepted where the momentum of the spectator proton was between
100 and 250 MeV/c \cite{Giacomelli2}. 
The
data disagree at forward angles by roughly a factor of two. 
Our model calculation for the $K^+d{\to}K^+np$ reaction with spectator 
proton (solid lines Fig.~\ref{kade9}) is in reasonable agreement with the 
differential cross sections measured by Giacomelli et al. \cite{Giacomelli2}. 
There is also a rough agreement with the data by Damerell et al. 
if one disregards the shaded area, except for the lowest kaon momentum
$p_K=434$ MeV/c.

Because most of the $K^+d{\to}K^+np$ differential cross sections were
measured only for limited scattering angles and/or there were 
ambiguities in the data analysis and, specifically, in the separation of
the break-up and coherent reactions, these data 
were never used to obtain integrated cross sections. 
Nonetheless, we present here predictions of our model for the integrated
$K^+d{\to}K^+np$ reaction cross section. Corresponding results
are shown in Fig.~\ref{kade7} (dashed-dotted line) as function of 
kaon momentum. We also indicate (by the open squares) the only 
published concrete data for the $K^+d{\to}K^+np$ integrated cross section 
\cite{Giacomelli3} that we found in the literature. These three points
are reasonably described by the model calculation.

\subsection{The reaction { ${\rm K^+d{\to}K^0pp}$}} 

\begin{figure}[t]
\vspace*{-6mm}
\centerline{\hspace*{3mm}\psfig{file=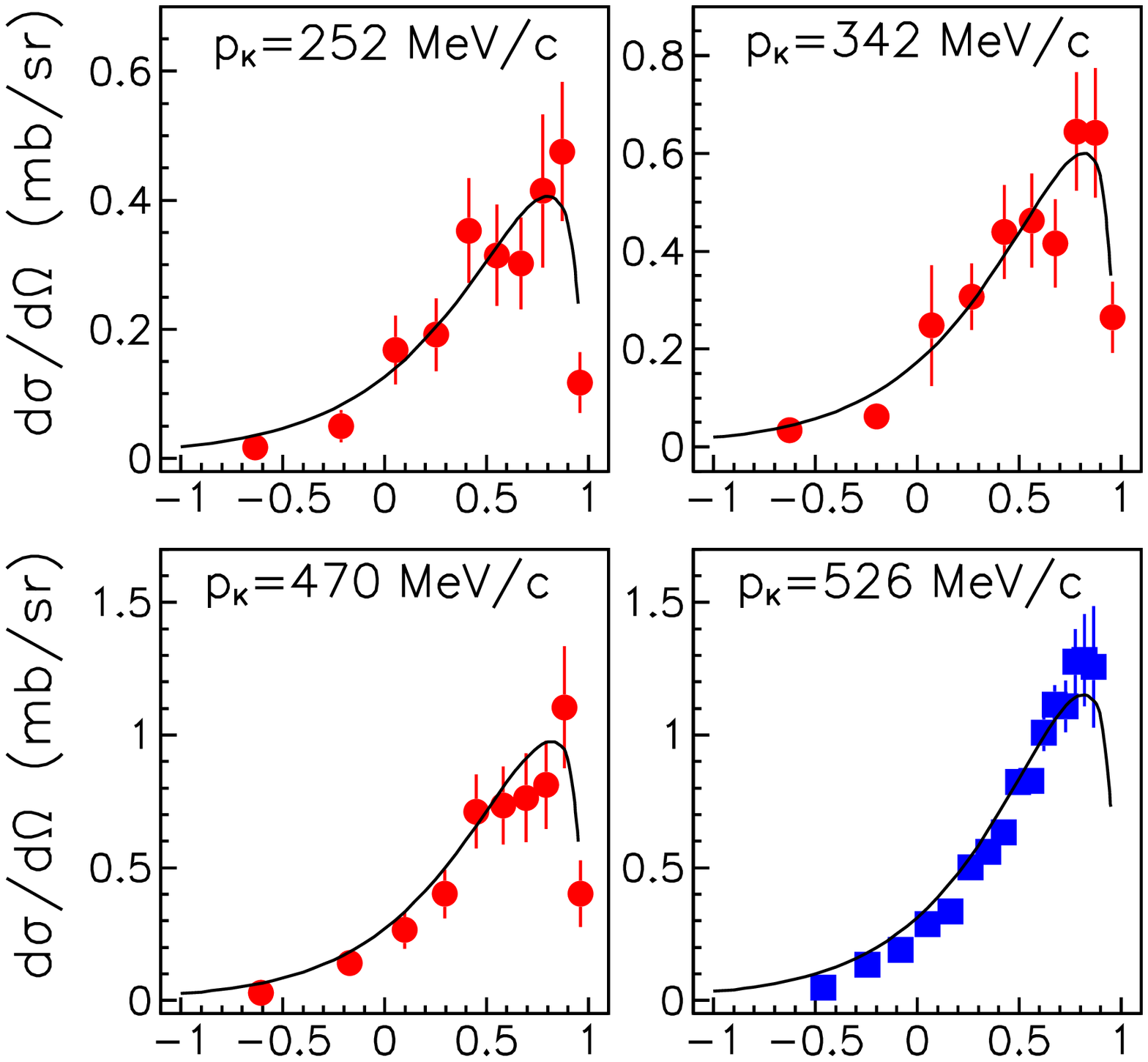,width=8.3cm,height=10.7cm}\hspace*{-15mm}\psfig{file=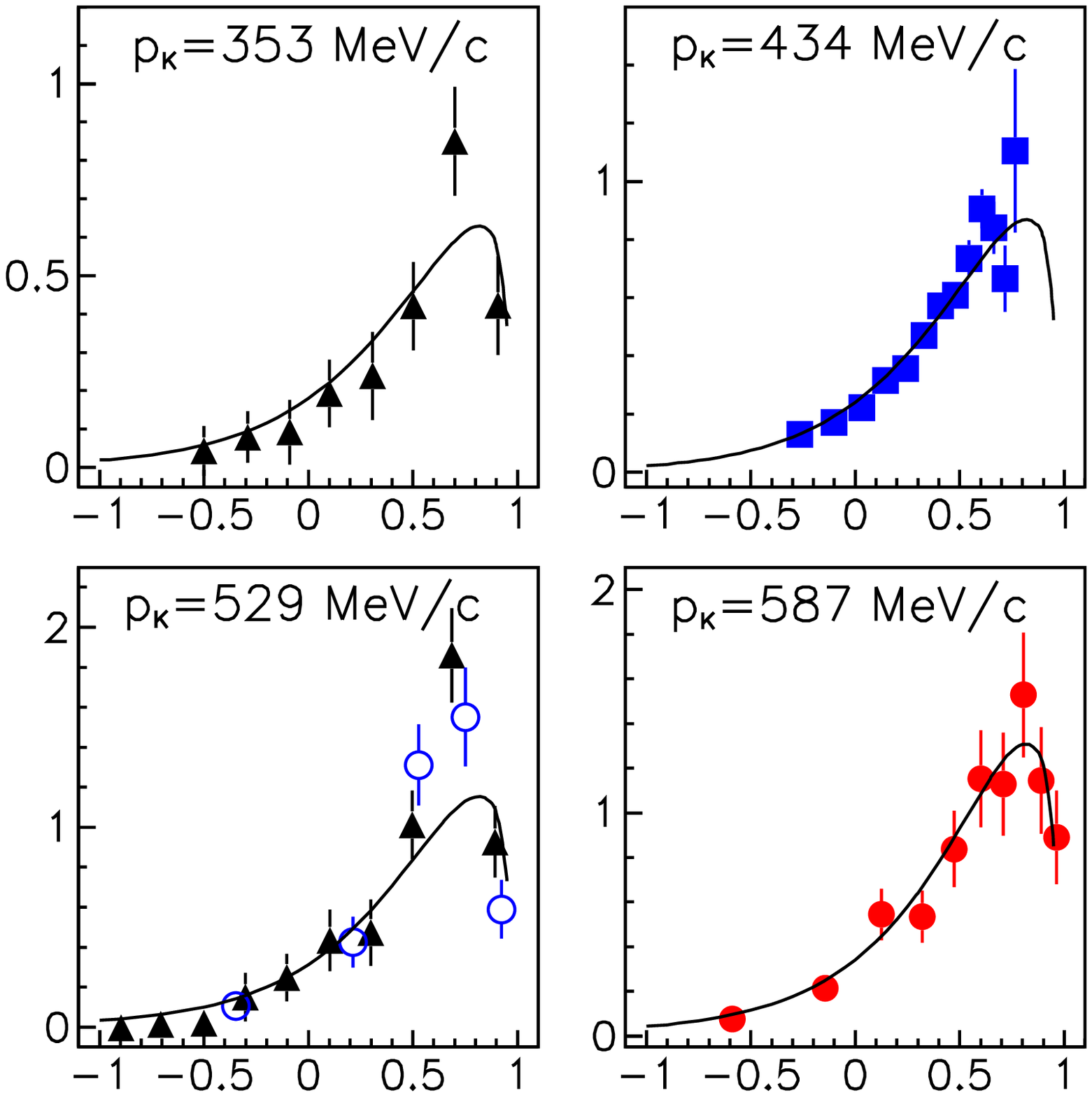,width=8.3cm,height=10.7cm}}
\vspace*{-17mm}
\centerline{\hspace*{3mm}\psfig{file=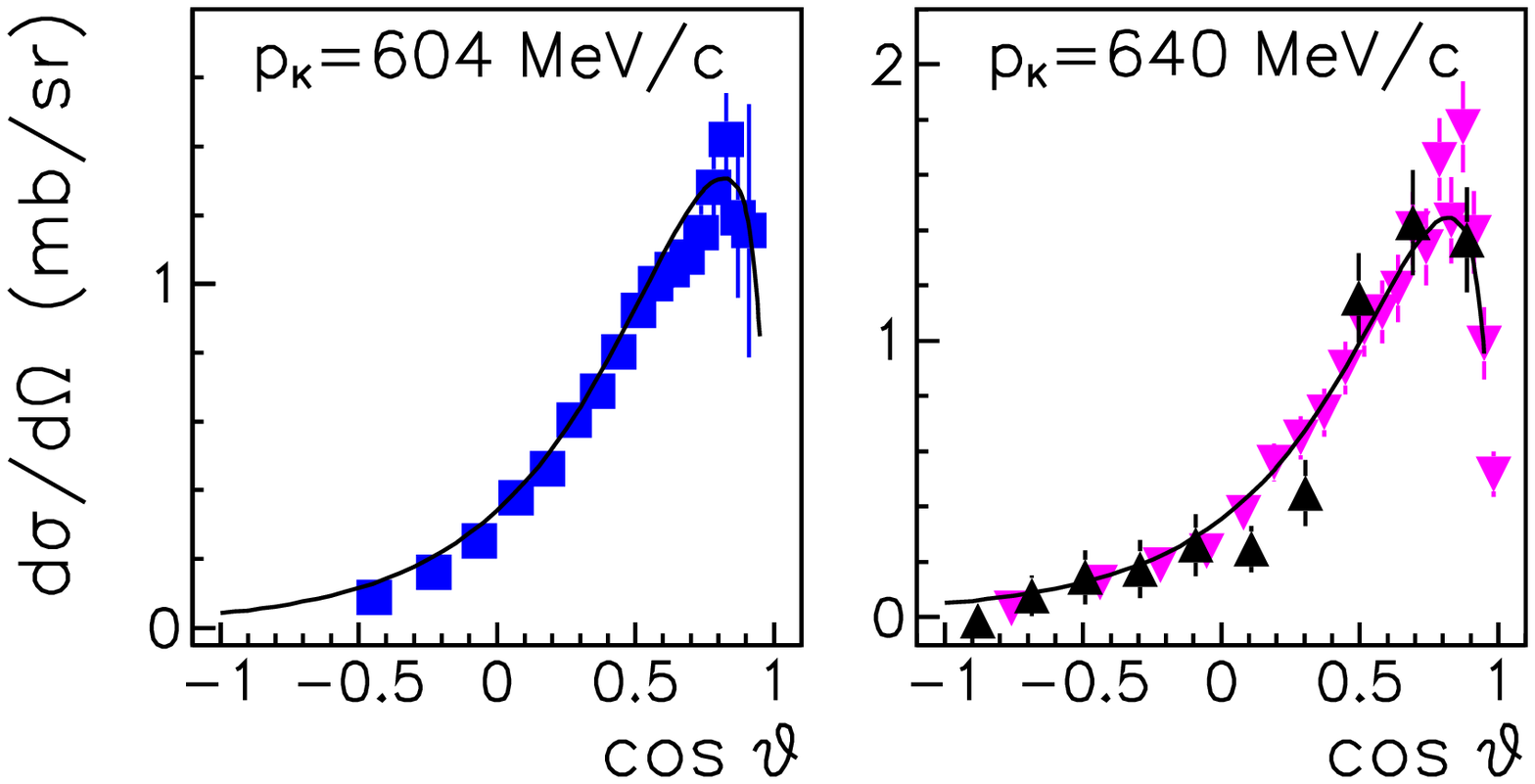,width=8.3cm,height=10.7cm}\hspace*{-15mm}\psfig{file=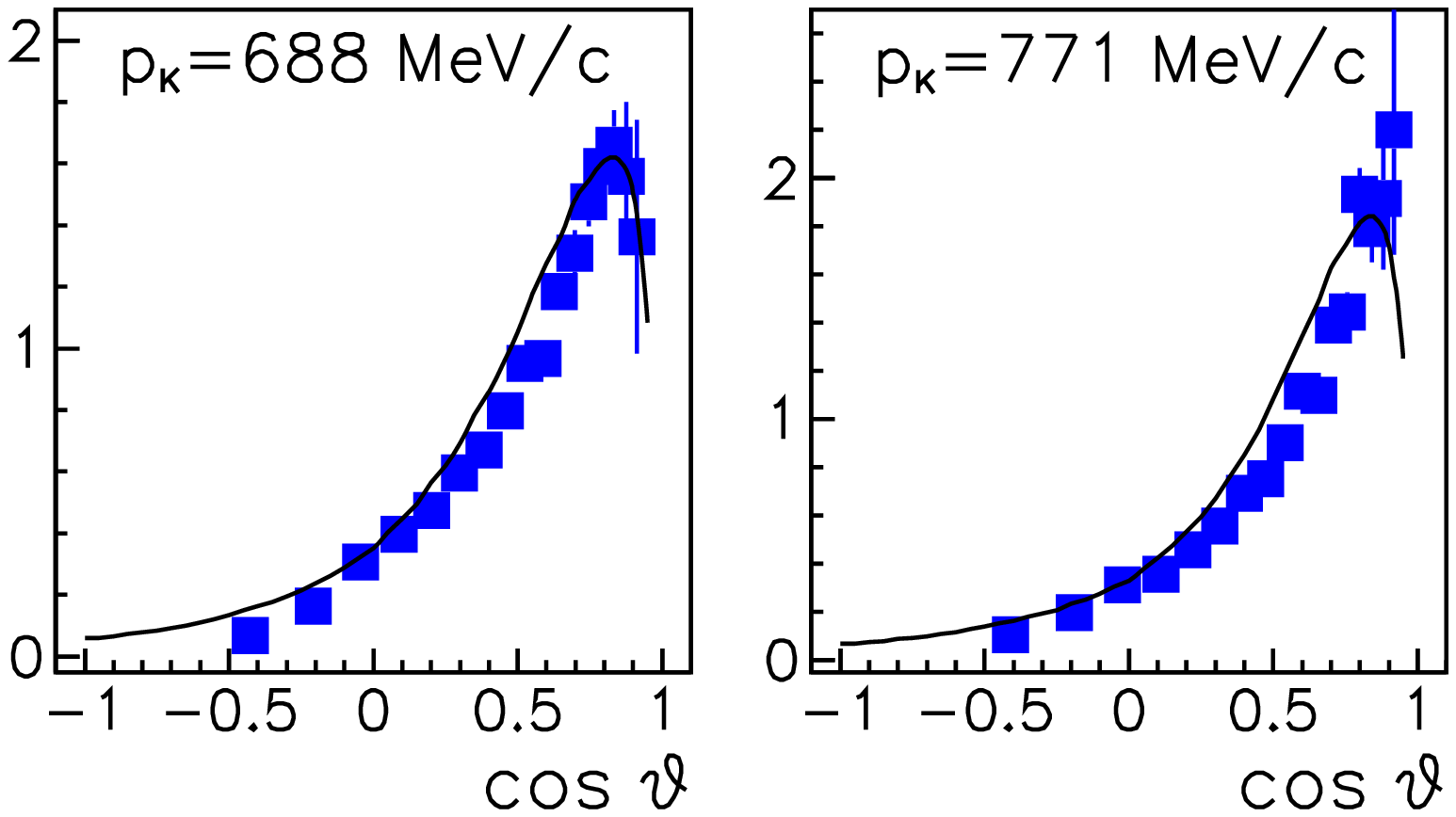,width=8.3cm,height=10.7cm}}
\vspace*{-50mm}
\caption{The $K^+d{\to}K^0pp$ differential cross section for different
incident kaon momenta as a function of the kaon scattering angle in the
laboratory system. The data are taken from Refs.~\cite{Stenger} (open circles) 
\cite{Glasser} (filled circles), \cite{Damerell} (squares), \cite{Giacomelli} 
(inverse triangles), and \cite{Slater} (triangles). 
The solid lines show our model calculations.} 
\label{kade16}
\end{figure}
The $K^+d{\to}K^0pp$ reaction can be uniquely reconstructed by
detecting two charged particles in the final state in addition to the
associated V track from the $K^0{\to}\pi^+\pi^-$ decay. As a consequence
the experimental error bars are smaller than those for the other 
break-up channels. An additional
advantage of the charge exchange reaction is due to the isospin structure,
cf. Eq.~(\ref{eq5}). The $K^+d{\to}K^0pp$ amplitude is proportional to the 
difference between the $I{=}1$ and $I{=}0$ amplitudes. 
On the other hand, since the $K^+d{\to}K^+pn$ reaction was measured without 
identification of the spectator proton the break-up amplitude amounts 
to the sum of 3/2 of the $I{=}1$ and only 1/2 of the $I{=}0$ amplitudes. 
Therefore the charge exchange $K^+d$ reaction is much more sensitive to 
the $I{=}0$ amplitude and, in turn, much more decisive for the
determination of the $I{=}0$ amplitude in a partial wave analysis.

A detailed comparison of our model calculation with the available experimental 
information for the $K^+d{\to}K^0pp$ charge exchange reaction at kaon momenta 
below 640~MeV/c was already presented in Ref.~\cite{Sibirtsev1}. 
For completeness we show those results again and we include predictions for 
two more momenta, namely 688 and 771 MeV/c, cf. Fig.~\ref{kade16}. 
Obviously, there is almost perfect agreement between the data and
the calculations, which might be not surprising since the J\"ulich model 
describes rather well the $I{=}0$ phase shifts, which were extracted from 
the charge exchange reaction. 
It is worthwile to note that the $K^+d{\to}K^0pp$ cross section 
at $p_K{=}252$~MeV/c constitutes the lowest energy at which data for $K^+d$ 
are available, and it can be well reproduced by the model calculation 
within the impulse approximation. 

Fig.~\ref{kade7} shows the integrated $K^+d{\to}K^0pp$ cross section as
a function of the kaon momentum. Also here there is nice agreement
of our calculation with the data, taken from Refs.~\cite{Damerell,Slater}.

\section{Coherent {\boldmath $K^+d$} scattering}

\subsection{Formalism}

The formalism for the coherent $K^+d{\to}K^+d$ scattering is given in
detail in Refs.~\cite{Hashimoto,Stenger,Sidhu,Giacomelli4,Bertocchi}.
Here we only list the formulas relevant for our calculation. 
Within the single scattering impulse approximation 
the elastic scattering amplitude is given by 
\begin{eqnarray}
A_d=\int d^3p\,\, \Psi ({\bf p})[A_p(k_0,k,p){+}A_n(k_0,k,p)]\,\,
\Psi({\bf p}{-}{\bf q}/2),
\label{coh1}
\end{eqnarray}
with ${\bf q}{=}{\bf k}_0{-}{\bf k}$ being the three-momentum transfered
from the initial to the final kaon. 
Assuming again that the $KN$ scattering amplitude is a 
smooth function of ${\bf p}$ as compared with the momentum dependence of
the deuteron wave function
the amplitude can be taken out of the integral in Eq.~(\ref{coh1}),
and the $KN$ amplitude can be taken approximately at 
${\bf p}{=}0$~\cite{Hashimoto,Bertocchi,Alberi}. Apparently, such an
approximation is even more justified for coherent scattering than 
for the break-up reactions. If one neglects also the deuteron $d$-wave
component then the coherent differential cross 
section is given by
\begin{eqnarray}
\frac{d\sigma}{d\Omega}=\left[|F_p+F_n|^2+\frac{2}{3}|G_p+G_n|^2\right]
\Phi_S^2(\frac{q}{2}),
\end{eqnarray}
where $\Phi_S$ is the spherical form factor of the deuteron 
evaluated from the $s$-wave deuteron wave function 
alone~\cite{Hashimoto,Stenger}. 

In some of the analyses~\cite{Glasser} the $d$-wave part of the 
deuteron wave function was taken into account
in order to estimate the single scattering contribution at large $q$ or
large scattering angles. In that case the coherent differential cross 
section is given by 
\begin{eqnarray}
\frac{d\sigma}{d\Omega}=|F_p+F_n|^2\left[\Phi_S^2(\frac{q}{2})
+\Phi_Q^2(\frac{q}{2})  \right]
+\frac{2}{3}|G_p+G_n|^2
\Phi_M^2(\frac{q}{2}),
\label{coh2}
\end{eqnarray}
where $\Phi_S$, $\Phi_Q$ and $\Phi_M$ are the spherical, quadrupole and
magnetic form factors of the deuteron, i.e. 
\begin{eqnarray}
\Phi_S&=&\Phi_a+\Phi_b, \nonumber \\
\Phi_Q&=&2\Phi_c-\frac{\Phi_d}{\sqrt{2}} \nonumber \\
\Phi_M&=&\Phi_a-\frac{\Phi_b}{2}+\frac{\Phi_c}{\sqrt{2}}
+\frac{\Phi_d}{2},
\end{eqnarray}
with 
\begin{eqnarray}
\Phi_a (q)&=&\int dr \,\, j_0(qr)\,\, |u(r)|^2, \nonumber \\
\Phi_b (q)&=&\int dr \,\, j_0(qr)\,\, |w(r)|^2, \nonumber \\
\Phi_c (q)&=&\int dr \,\, j_2(qr)\,\, u(r)\,w(r), \nonumber \\
\Phi_d (q)&=&\int dr \,\, j_2(qr)\,\, |w(r)|^2, \nonumber \\
\end{eqnarray}
where $j_0$ and $j_2$ are the zeroth and second order spherical Bessel 
functions, respectively. 
In general one should take into account all these form
factors in analysing scattering at large angles and for evaluating
the contribution from multiple scattering. That is why we
provide the single scattering formalism here explicitely. 
As before we utilize the deuteron wave functions 
of the CD Bonn potential~\cite{Machleidt}.

\subsection{Results}

The $K^+d$ coherent scattering at kaon momenta below 800 MeV/c
was measured in several experiments~\cite{Glasser,Giacomelli4,Sakitt}.
It is clear from Eqs.~(\ref{coh1}) and (\ref{coh2}) that the angular
distribution for coherent scattering is dominated by the deuteron form 
factor and the $I{=}1$
component of the $K^+N$ scattering amplitude. That actually leads to
the conclusion~\cite{Glasser} that the $K^+d{\to}K^+d$ reaction should
not be too sensitive to the $I{=}0$ amplitude, or in other words, 
one has to expect large uncertainties when one tries to extract the
$I{=}0$ amplitude from $K^+d$ coherent scattering.

\begin{figure}[t]
\vspace*{-6mm}
\centerline{\hspace*{3mm}\psfig{file=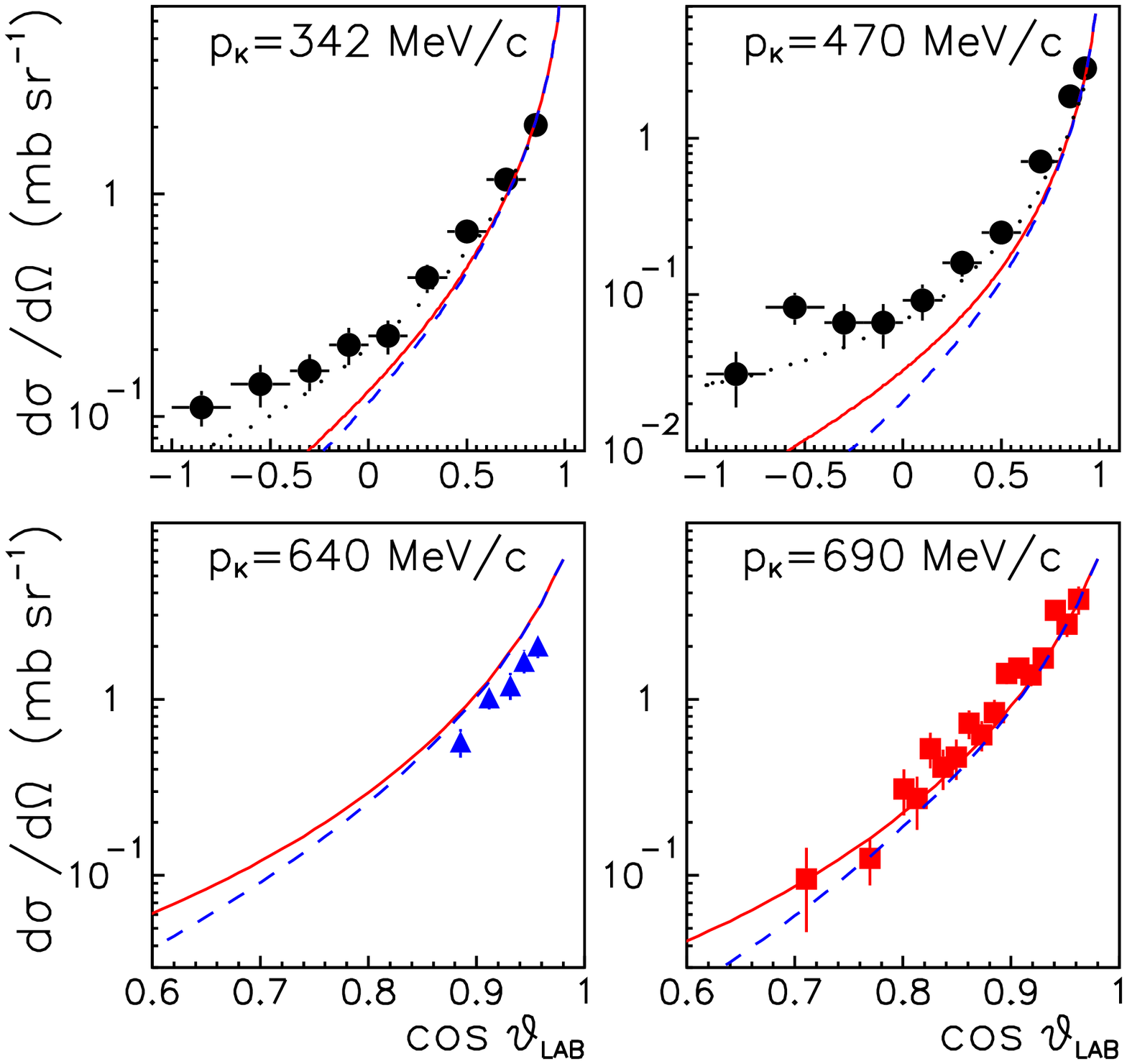,width=7.9cm,height=10.7cm}\hspace*{-9mm}\psfig{file=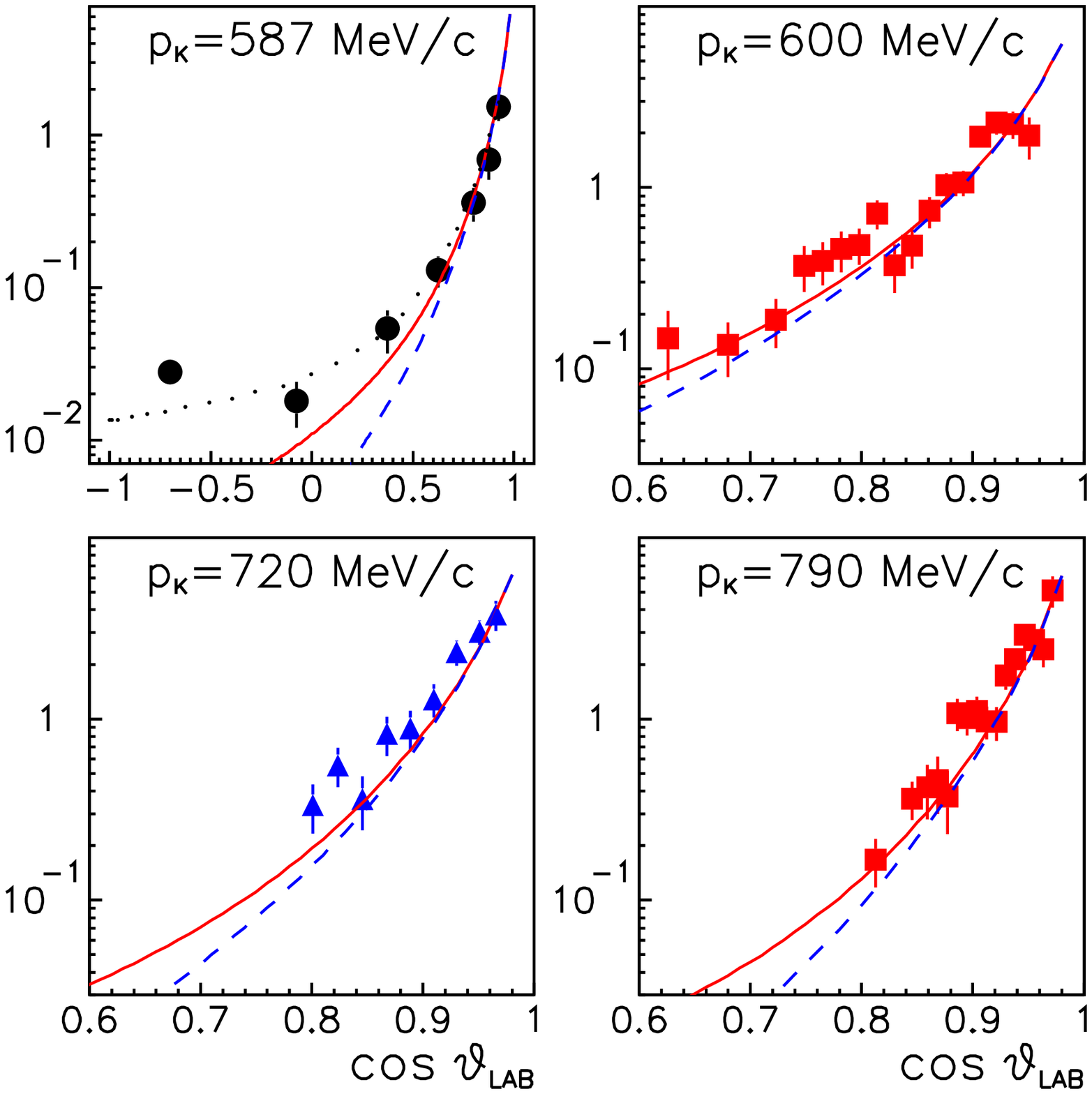,width=7.9cm,height=10.7cm}}
\vspace*{-5mm}
\caption{The $K^+d{\to}K^+d$ differential cross section for different
kaon momenta as a function of the kaon scattering angle in the
laboratory system. The data are taken from Ref.~\cite{Glasser} (circles),
\cite{Giacomelli4} (triangles), and \cite{Sakitt} (squares). 
The solid lines show our calculations by Eq.~(\ref{coh2}), while the dashed 
lines are results obtained with Eq.~(\ref{coh1}).
The dotted lines show the relativistic Faddeev
calculations~\cite{Garcilazo}.} 
\label{penco2}
\end{figure}

In Fig.~\ref{penco2} we compare our model results with the 
available data on the differential $K^+d{\to}K^+d$ cross section.
The measurement of Glasser et al.~\cite{Glasser} (filled circles)
was performed at the kaon momenta of 342, 470, and 587 MeV/c.
The solid lines show our calculations by Eq.~(\ref{coh2}), while the
dashed lines are results from Eq.~(\ref{coh1}). While the data at forward
angles are well reproduced we find a strong discrepancy at large
angles. This is not too surprising because it is known from Faddeev 
calculations of the $K^+d$ system that multiple-scattering effects 
play an important role at backward angles in elastic 
$K^+d$ scattering \cite{Sanudo,Garcilazo}. Specifically, the 
corresponding results presented in Ref.~\cite{Garcilazo} demonstrate 
very clearly that the description of the data of Glasser et al. improves 
significantly in this angular region as compared to the impulse
approximation. For the ease of comparision we included the curves 
of Ref.~\cite{Garcilazo} in our Fig.~\ref{penco2} (dotted lines). 
At the same time those Faddeev calculations confirm that the 
impulse approximation works rather well in forward direction, i.e.
for laboratory angles smaller than 60 degrees 
($\cos \theta > 0.5$), say. 
Incidentally, since this angular range provides the bulk contribution to
the integrated elastic $K^+d$ cross section one expects that the 
impulse approximation should yield also reliable results for the
latter observable. This is indeed the case, judging from the results
presented in Fig. 1 of Ref.~\cite{Garcilazo}, where one sees 
practically no difference to the Faddeev calculation even for rather
small energies. 
 
As a side remark we would like to point out that none of the calculations 
can reproduce an apparent structure visible in the tail of the coherent 
angular spectrum at kaon momentum of 470 MeV/c. But it is possible that
this structure is simply a statistical fluctuation in the data. 
 
Let us mention in this context that the authors of Ref.~\cite{Glasser} 
were able to reproduce the angular spectra both at small and large angles 
within the impulse approximation by suitably adjusting the $I=0$ $KN$ 
PW amplitudes to the $K^+d{\to}K^+d$ reaction, which resulted in
large $p$-wave contributions. However, in that case the obtained 
PW amplitudes are in strong conflict with the solution from 
their own analysis of the $K^+d{\to}K^0pp$ reaction~\cite{Glasser}. 

The squares in Fig.~\ref{penco2} show the data of Sakitt et
al. \cite{Sakitt}. Their measurement covers only forward angles and it is
well described by our calculations. The same is the case for the 
results by Giacomelli et al. \cite{Giacomelli4}, which are shown
by triangles in the Fig.~\ref{penco2}. 

\begin{figure}[t]
\vspace*{-6mm}
\centerline{\hspace*{3mm}\psfig{file=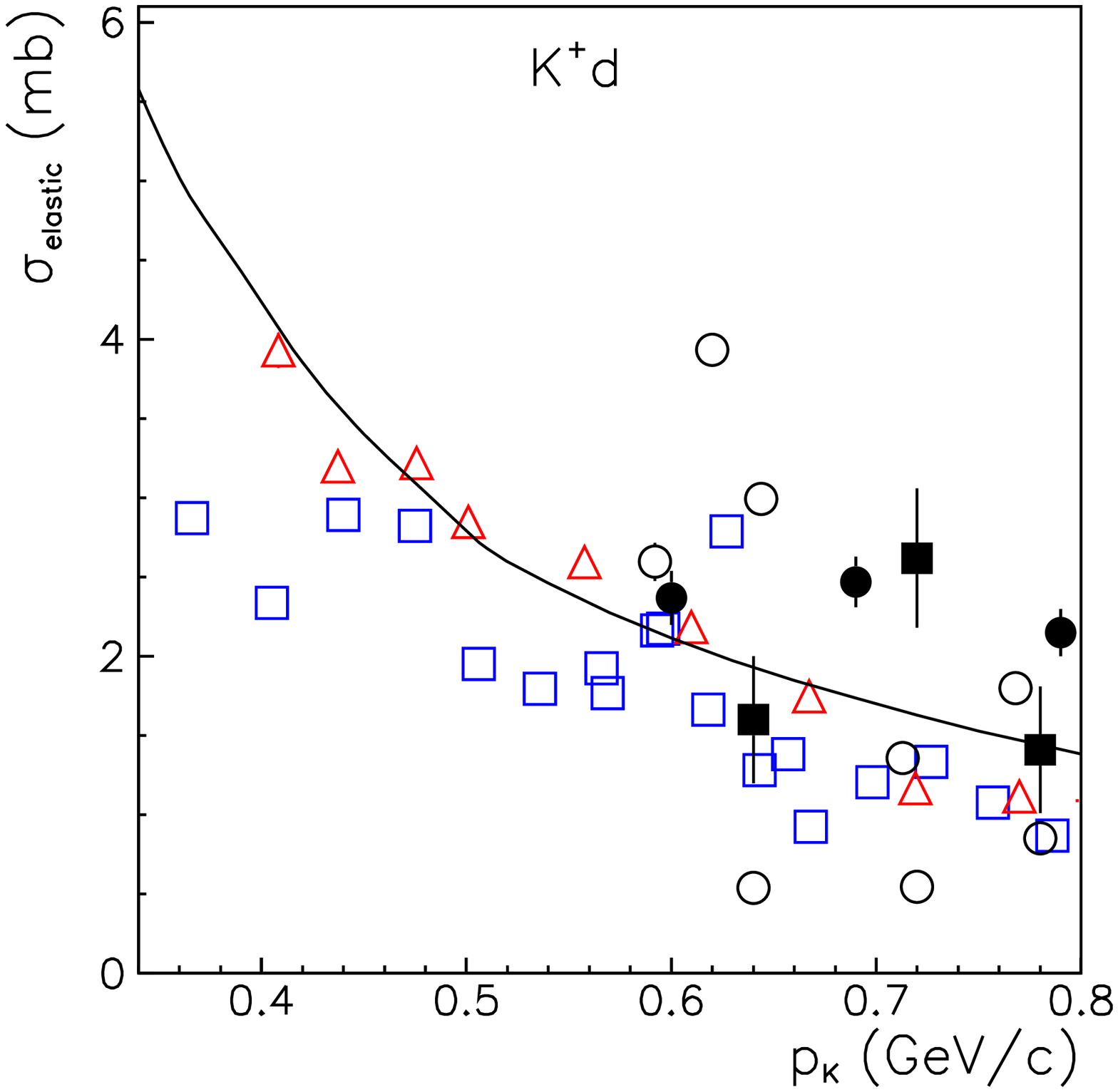,width=8.cm,height=9cm}\hspace*{-3mm}\psfig{file=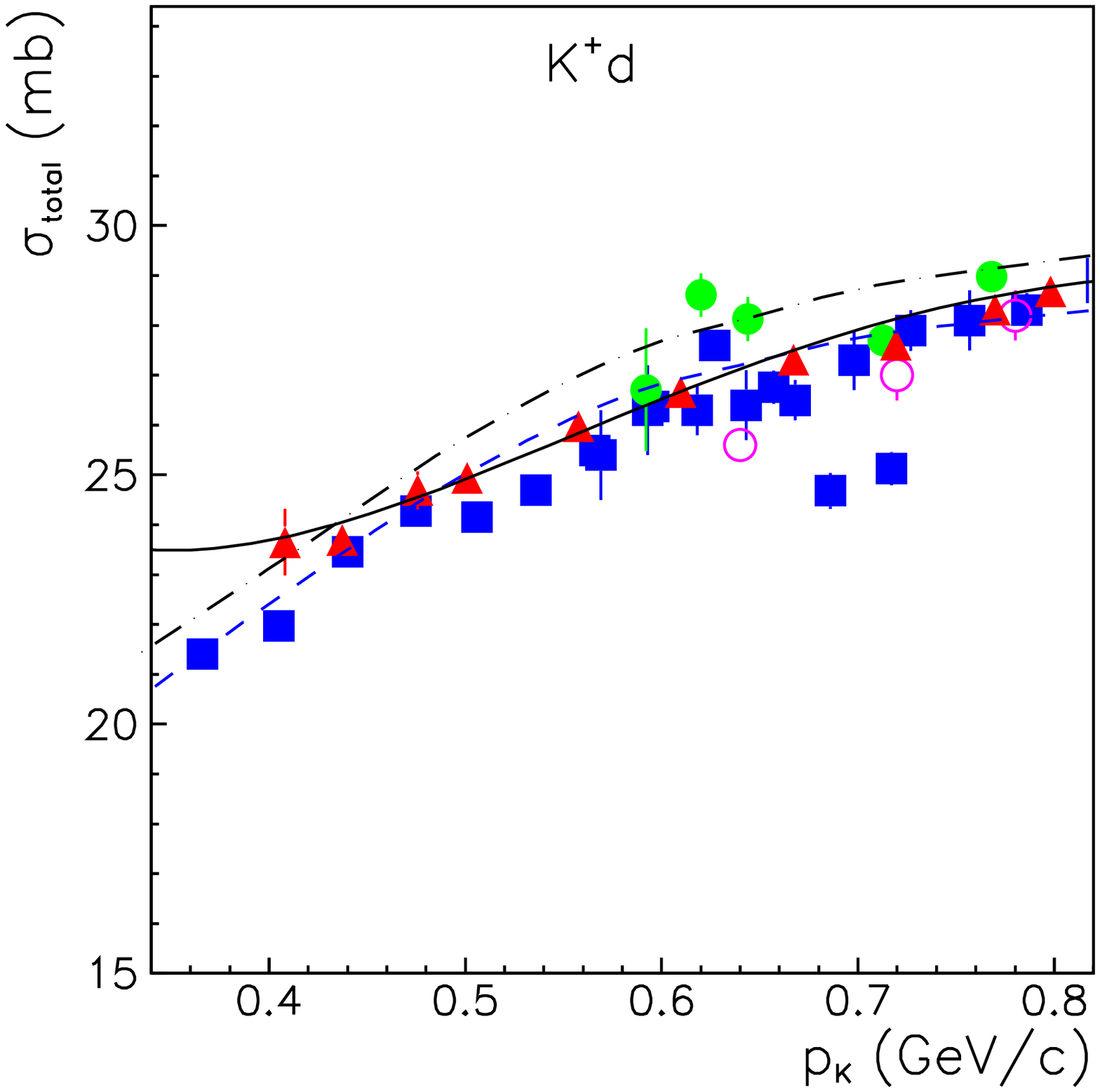,width=8.cm,height=9cm}}
\vspace*{-5mm}
\caption{The $K^+d$ elastic (left) and total (right) cross section as a function of the
kaon momentum. The data for the $K^+d$ total cross section are 
taken from Refs.~\cite{Carroll,Bowen1,Bowen2,Bugg1,Giacomelli3}.
Available data on the integrated $K^+d$ elastic cross section 
from direct measurments are given by filled symbols \cite{Giacomelli4,Sakitt}.
The open symbols represent values obtained by subtracting our model results 
for the integrated $K^+d{\to}K^+np$ plus $K^+d{\to}K^0pp$ cross sections 
from the various data on the $K^+d$ total cross section. 
The solid lines represent our model calculation, where the total cross 
section is simply the sum of all partial cross sections. 
The dash-dotted line is the result of Eq.~(\ref{glaub}) with
$\delta\sigma$=0, while the dashed line is obtained with a cross section 
defect evaluated via Eq.~(\ref{corr1}) with the ratios $\rho_p$ and
$\rho_n$ taken from the J\"ulich $KN$ model.
} 
\label{kade12}
\end{figure}

Results for the integrated $K^+d$ cross sections are presented in
Fig.~\ref{kade12} as a function of the kaon momentum. In the left
panel we show results for the elastic $K^+d{\to}K^+d$ scattering
cross section. Here the closed symbols are data from direct
measurements while the open symbols 
represent values obtained by subtracting our model results 
for the integrated $K^+d{\to}K^+np$ plus $K^+d{\to}K^0pp$ cross sections 
from the various data on the $K^+d$ total cross section. These (deduced) 
elastic cross sections fluctuate substantially. It is simply a
reflection of the discrepancy between these data for the total 
$K^+d$ cross sections, specifically in case of 
Bowen et al.~\cite{Bowen1,Bowen2} and Carroll et al.~\cite{Carroll}. 
The solid line is our prediction for the elastic cross section which 
is roughly in line with the directly measured and the deduced data. 
As already mentioned above, we expect that the impulse approximation
leads to reliable results over the whole considered momentum range. 

\section{The total {\boldmath ${\rm K^+d}$} cross section}

The total $K^+d$ cross section is presented in the right panel of 
Fig.~\ref{kade12}. The data are the same as in the Fig.~\ref{kade7}. 
The solid line represents the sum of the calculated
$K^+d{\to}K^+np$, $K^+d{\to}K^0pp$ and the elastic $K^+d$ 
cross sections. Our result is in good agreement with the data for 
kaon momenta $p_K \ge 0.45$ GeV/c. This is not too surprising since our 
model calculation yields reasonable descriptions of the differential 
cross sections in all contributing individual reaction channels.

In the extraction of the $I{=}0$ isospin cross section (cf. Fig.~\ref{kade11})
from $K^+d$ experiments~\cite{Carroll,Bowen1,Bowen2,Cool}
the total $K^+d$ cross section is written as~\cite{Glauber,Franco}
\begin{eqnarray}
\sigma_{K^+d}^{tot}=\sigma_{K^+p}^{tot}+\sigma_{K^+n}^{tot}-\delta\sigma,
\label{glaub}
\end{eqnarray}
where $\sigma_{K^+p}^{tot}$ and $\sigma_{K^+n}^{tot}$ are the total 
$K^+p$ and $K^+n$ cross sections (cf. Fig.~\ref{kade11}) and
$\delta\sigma$ is the so-called cross section defect. The first two 
terms on the right hand side of Eq.~(\ref{glaub}) follow from the
application of the optical theorem to the $K^+d$ amplitude within
the impulse approximation. Specifically, 
when taking the $K^+N$ scattering amplitude out of the $p$ 
integral in Eq.~(\ref{coh1}), one ends up with an 
integration over the deuteron wave function. This yields the deuteron
form factor which is one at ${\bf q}{=}0$ by definition, and thus 
$A_d(0){=}A_p(0){+}A_n(0)$. Applying then the optical theorem to $A_d$, 
\begin{eqnarray}
\sigma_{tot}=\frac{4\pi}{p_K}\,{\rm Im}\, A_d(0) \ , 
\end{eqnarray}
one obtains the first two terms. The third term, $\delta\sigma$, 
is usually derived from the double scattering amplitude according 
to the Glauber theory~\cite{Glauber,Franco}, which should be valid
for high energies. At low energies the term should contain 
additional corrections due to multiple scattering.

To inspect how large this correction is one can assume 
$\delta\sigma${=}0 and compare the outcome of Eq.~(\ref{glaub}) 
with the data. The corresponding result is shown by the dash-dotted line 
in Fig.~\ref{kade12}. We see that the cross section defect amounts
to about 1 mb or to about 4\% of the total $K^+d$ cross section.

In all experiments~\cite{Carroll,Bowen1,Bowen2,Bugg1,Cool,Giacomelli3} 
devoted to the extraction of the total $I{=}0$ isospin
$K^+N$ cross section from the measured total $K^+d$ 
cross section Eq.~(\ref{glaub}) was used with
$\delta\sigma$ explicitely given by
\begin{eqnarray}
\delta\sigma{=}\frac{\langle r^{-2}\rangle}{4\pi}[2\sigma_{K^+p}\sigma_{K^+n}
(1{-}\rho_p\rho_n){-}
\frac{1}{2}\sigma_{K^+p}(1{-}\rho_p)^2{-}
\frac{1}{2}\sigma_{K^+n}(1{-}\rho_n)^2,
\label{corr1}
\end{eqnarray}
where $\rho_p$ and $\rho_n$ are the ratios of the real to imaginary
part of the kaon scattering amplitude on proton and neutron,
respectively, while $\langle r^{-2}\rangle{=}0.3$ fm$^{-2}$ is the parameter of
the Glauber model introduced during the evaluation of the double
scattering amplitude and corresponds to the averaged distance between
proton and neutron in the deuteron.
The correction according to Eq.~(\ref{corr1}) includes charge
exchange~\cite{Wilkin} in the double scattering. 
At high energies the scattering
amplitude are almost imaginary and $\rho_p$=$\rho_n$=0. Therefore,
$\delta\sigma$ is given in terms of the $K^+p$ and $K^+n$ cross
section. Since $\sigma_{K^+p}^{tot}$ is experimentaly  known one can
easily extract $\sigma_{K^+n}^{tot}$ from  $\sigma_{K^+d}^{tot}$.

\begin{figure}[t]
\vspace*{-6mm}
\centerline{\hspace*{3mm}\psfig{file=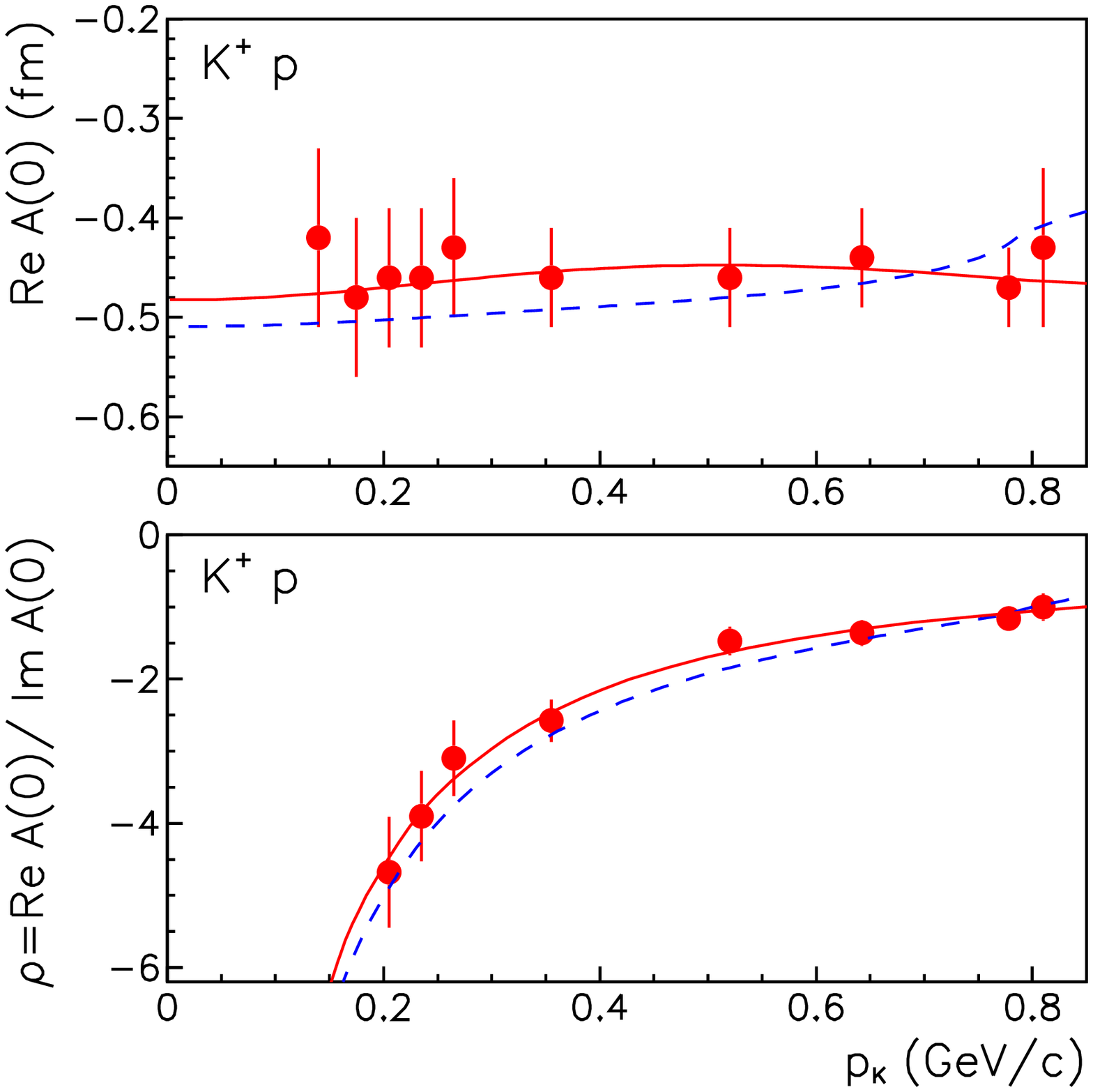,width=8.cm,height=9.4cm}\hspace*{-13mm}\psfig{file=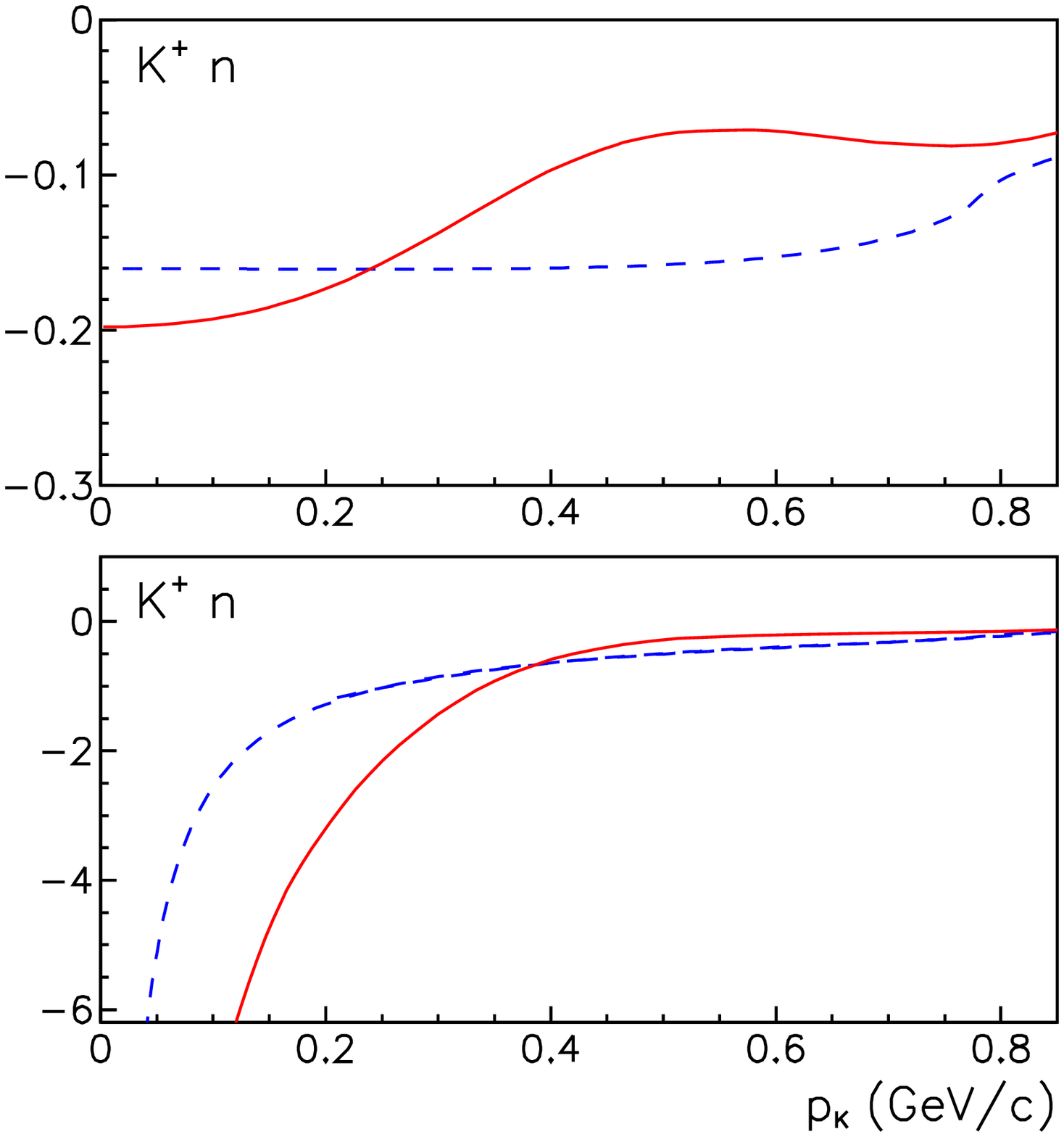,width=8.cm,height=9.4cm}}
\vspace*{-5mm}
\caption{The real parts of the $K^+p$ and $K^+n$ forward scattering
amplitudes in the laboratory system and the 
ratios $\rho$ of the real to imaginary parts as a
function of the kaon momentum. The data were taken from
Ref.~\cite{Dumbrais}. The solid lines are the predictions 
of the J\"ulich model I, while the dashed lines show the results from 
dispersion relations~\cite{Sibirtsev5}.} 
\label{kade14}
\end{figure}

For very low momenta the imaginary part of the $KN$ forward 
scattering amplitude approaches zero, while the real part does not. 
Furthermore, the real part of the scattering amplitude in
forward direction, ${\rm Re}\,{A(0)}$, for $K^+p$ scattering is known 
experimentally. Corresponding results~\cite{Dumbrais}
are shown in Fig.~\ref{kade14} together with the ratio $\rho_p$. 
Note that there is no
experimental information on ${\rm Re}\,{A(0)}$ for $K^+n$ scattering. 
The solid lines in Fig.~\ref{kade14} show the predictions of the 
J\"ulich $KN$ model, while the dashed lines show results of a 
calculation by dispersion relations~\cite{Sibirtsev5}. For $K^+p$ 
scattering both results are similar and also in agreement with the 
data. However, not unexpectedly, there is a 
substantial disagreement between the dispersion calculation and the 
J\"ulich result for the forward $K^+n$ scattering amplitude,
in particular for small momenta. Presumably the variations of the $I{=}0$ 
cross section in Fig.~\ref{kade11} between different experiments are due 
to different parameters used in the application of Eq.~(\ref{corr1}). 

The
result for the total $K^+d$ cross section based on Eq.~(\ref{glaub}) with
$\delta\sigma$ evaluated via Eq.~(\ref{corr1}) for the $KN$ amplitude of
the J\"ulich model is presented by the dashed line in Fig.~\ref{kade12}. 
It is in remarkable qualitative agreement with the data over the whole
considered momentum range. 
 
\section{Summary}

We investigated the $K^+d$ reaction at kaon momenta below 800
MeV/c. The available data on the channels $K^+d{\to}K^0pp$, $K^+d{\to}K^+pn$ 
and $K^+d{\to}K^+d$ and the total $K^+d$ reaction cross section
were compared with a calculation based on the J\"ulich $KN$ model. 
The angular spectra were computed within the
single scattering impulse approximation taking into account Fermi
motion of the nucleons in the deuteron and the final three-body
kinematics for the charge exchange and break-up reactions.
A compact summary of the considered data and the achieved results
is given in Table \ref{data}. 

It was found that the experimental results for the reaction $K^+d{\to}K^0pp$
published in Refs.~\cite{Stenger,Glasser,Damerell,Giacomelli,Slater}
can be very well reproduced down to the kaon momentum of 252~MeV/c, 
i.e. even for the smallest momentum where data are available. 
There seems to be practically no space for additional effects with
regard to possible corrections to the impulse approximation, although 
one would expect that such corrections might become more and more 
relevant when approaching the threshold. 
It is worthwile to notice the agreement between the data and the
calculations, despite the fact that in the experiments the momentum 
of the spectator proton was not detected and no additional cuts were 
imposed. While the shape of the momentum spectrum of the spectator 
proton allows to examine the applicability of the single scattering 
impulse approximation~\cite{Alberi}, a momentum cut permits for the
isolation of the multiple scattering contribution. These
conditions were neither monitored nor imposed in the experiments.

The reaction $K^+d{\to}K^0pp$ constrains to a large extent the
$I{=}0$ $KN$ scattering amplitude because of two reasons.
First, it can be uniquely identified by the two
charged particles in the final state associated with a $V$ track from the
$K^0{\to}\pi^+\pi^-$ decay and therefore the data are relatively
precise. Second, the charge exchange amplitude is
the half difference between the $I{=}1$ and $I{=}0$ amplitudes,
i.e. it contains a sizeable isoscalar contribution. 
Note that the $I{=}1$ amplitude can be uniquely determined from the 
$K^+p{\to}K^+p$ data.
 
The data on the reaction $K^+d{\to}K^+pn$ turned out to be practically 
unimportant for the $KN$ phase-shift analyses although there is a 
large amount of experimental results for this reaction. 
This has to do with the fact that, in general the spectator 
nucleon from the $K^+d{\to}K^+pn$ reaction was not measured and therefore
the amplitude for the break-up reaction is the sum of the $K^+p{\to}K^+p$
and $K^+n{\to}K^+n$ amplitudes. In that case the $K^+d{\to}K^+pn$
amplitude is dominated by the $I{=}1$ component, i.e. its contribution
constitutes 3/4 of the total reaction amplitude while only 1/4 come from 
the $I{=}0$ component. 
Moreover, there are also difficulties in the final particle identification, 
especially in the $K^+$-meson forward direction, where the various 
experimental groups have tried to resolve that problem in different ways. 
As a consequence in this angular range the results from different measurements 
are partly in contradiction with each others. 
It turned out that 
our model calculation is in nice agreement with the $K^+d{\to}K^+pn$ data 
from Refs.~\cite{Stenger,Damerell}, but it fails to reproduce the results 
from Ref.~\cite{Glasser} at forward angles. 

We also analysed those experiments~\cite{Damerell,Giacomelli2} which 
aimed to identify the spectator nucleon and to measure the
elementary scattering using the (other) nucleon in the deuteron. 
Although we reasonably reproduce the $K^+n{\to}K^+n$ results
from Ref.~\cite{Giacomelli2}, our calculations are partly in strong
disagreement with data from Ref.~\cite{Damerell}. We should say that in
the latter experiment the spectator nucleon was 
not measured but indirectly reconstructed applying a somewhat obscure 
event selection procedure and, in addition, the authors admit that
there are extremely large uncertainties of the data for forward angles. 
In view of that it might be not too surprising that there are
discrepancies. 

One of the goals of the study in Ref.~\cite{Damerell} was to compare the
$K^+p{\to}K^+p$ results obtained with a deuteron target with those obtained 
on a free proton~\cite{Adams} and to verify the spectator model formalism. 
The obtained experimental results on a bound proton are partly in strong 
disagreement with those from free protons, depending on the angular range. 
However, we attribute this discrepancy to the just mentioned uncertainties
in the experimental spectator identification method and not to the validity 
of the spectator formalism. 

Finally we analysed $K^+d{\to}K^+d$ elastic scattering. Here 
our calculations within the impulse approximation reproduce
the data~\cite{Glasser,Giacomelli4,Sakitt} quite well at forward angles 
but exhibit substantial deviations at backward angles and at kaon
momenta below 600~MeV/c, say. Although this discrepancy accounts 
for not more that 10\% of the total elastic $K^+d{\to}K^+d$ cross section 
it explicitly indicates that strong few-body effect play a role in this 
specific kinematical regime. Indeed, relativistic $K^+d$ Faddeev calculations 
performed by Garcilazo \cite{Garcilazo} can reproduce the $K^+d{\to}K^+d$ 
scattering fairly well also at large angles.

Combining our results for the $K^+d{\to}K^0pp$, $K^+d{\to}K^+pn$ and
$K^+d{\to}K^+d$ reaction channels we computed the total $K^+d$ 
reaction cross section which turned out to be in good agreement with 
the measurements given in Refs.~\cite{Carroll,Bowen1,Bowen2,Bugg1,Giacomelli3}.

In conclusion, the bulk of the available $K^+d$ data 
base for $K^+$-meson momenta below 800~MeV/c, which comprises 
differential and integrated cross sections for the reaction
channels $K^+d{\to}K^0pp$, $K^+d{\to}K^+pn$, and $K^+d{\to}K^+d$, 
can be described quantitatively and consistently within
the single scattering impulse approximation utilizing a $KN$ model 
that reproduces the results of up-to-date partial-wave analyses.
This means that the data are indeed consistent with each other
(save a few exceptions discussed above) where one has to emphasize
that many observables were measured by three or more 
independent groups at various beam momenta. 
The success of the single scattering impulse approximation
also implies that for the kinematics where the data are available
multiple scattering effects are presumably negligible. In fact, 
the only clear evidence for the presence of such effects were
found in the $K^+d$ elastic cross section at backward angles. 
Finally, the nice reproduction of the $K^+d$ data based on a standard
$KN$ model leaves also little room for contributions of a 
$\Theta^+$(1540) pentaquark. There is no obvious signal in the data
that such a resonance is needed. On the other hand, a $\Theta^+$(1540)
with an rather small width of 1 MeV or less can always be accommodated,
as we have already shown in Ref.~\cite{Sibirtsev1}. This conclusion is in 
agreement with results of the most recent searches for the $\Theta^+$(1540) 
in the reaction $\gamma p \to \bar K^0K^+n$ and $\gamma p \to \bar K^0K^0p$ 
\cite{Battaglieri,De_Vita}.
 
\begin{table}[ht]
\caption{$K^+d$ data analyzed in the present study. Note that the results
for the reaction $K^+n\to K^+n$ are not directly from a measurement but
were extracted from the reaction $K^+d \to K^+np$ as described in the
text. For the quality rating we use the following categories: 
excellent agreement of data and theory (***), minor deviation (**), major 
deviation (*), where in the latter two cases we provide a concrete
description of the deficiency in a footnote. 
}
\label{data}
\begin{center}
\begin{tabular}{lcccl}
\hline
 & Ref. & $p_K$ [MeV/c] & type & quality \\
\hline
\hline
\multicolumn{5}{c}{$K^+d \to K^+np$} \\
\hline
\hline
Stenger &  \cite{Stenger}  & 377, 530 & $d\sigma/d\Omega$  & *** \\
Glasser &  \cite{Glasser}  & 342 - 587 & $d\sigma/d\Omega$ & * $^{\dagger)}$ \\
Damerell &  \cite{Damerell} & 434 - 771 & $d\sigma/d\Omega$ & *** \\
Giacomelli & \cite{Giacomelli3} & 640 - 780 & $\sigma_{total}$ & *** \\
\hline
\hline
\multicolumn{5}{c}{$K^+d \to K^0pp$} \\
\hline
\hline
Stenger & \cite{Stenger}  & 530 & $d\sigma/d\Omega$ & ** $^{\dagger\dagger)}$\\
Glasser & \cite{Glasser}  & 252 - 587 & $d\sigma/d\Omega$ & *** \\
Damerell & \cite{Damerell} & 434 - 771 & $d\sigma/d\Omega$ & *** \\
Giacomelli & \cite{Giacomelli} & 640 & $d\sigma/d\Omega$ & *** \\
Slater & \cite{Slater} & 353 - 640 & $d\sigma/d\Omega$ & ** $^{\dagger\dagger)}$\\
Damerell & \cite{Damerell} & 434 - 771 & $\sigma_{total}$ & *** \\
Slater & \cite{Slater} & 252 - 640 & $\sigma_{total}$ & *** \\
\hline
\hline
\multicolumn{5}{c}{$K^+d \to K^+d$} \\
\hline
\hline
Glasser & \cite{Glasser}  & 342 - 587 & $d\sigma/d\Omega$ & ** $^{\ddag)}$ \\
Giacomelli &  \cite{Giacomelli4} & 640, 720 & $d\sigma/d\Omega$ & *** \\
Sakitt & \cite{Sakitt} & 600 - 790 & $d\sigma/d\Omega$ & *** \\
Giacomelli &  \cite{Giacomelli4} & 640 - 780 & $\sigma_{elastic}$ & * $^{\ddag\ddag)}$ \\
Sakitt & \cite{Sakitt} & 600 - 790 & $\sigma_{elastic}$ & * $^{\ddag\ddag)}$ \\
Carroll & \cite{Carroll} & 408 - 798 & $\sigma_{total}$ & ** $^{\S)}$ \\
Bowen & \cite{Bowen1} & 366 - 717 & $\sigma_{total}$ & ** $^{\S)}$ \\
Bowen & \cite{Bowen2} & 569 - 786 & $\sigma_{total}$ & ** $^{\S)}$ \\
Bugg & \cite{Bugg1} & 592 - 768 & $\sigma_{total}$ & ** $^{\S)}$ \\
Giacomelli & \cite{Giacomelli3} & 640 - 780 & $\sigma_{total}$ & ** $^{\S)}$ \\
\hline
\hline
\multicolumn{5}{c}{$K^+n \to K^+n$} \\
\hline
\hline
Damerell & \cite{Damerell} & 434 - 771 & $d\sigma/d\Omega$  & * $^{\S\S)}$ \\
Giacomelli &  \cite{Giacomelli2} & 640 - 780 & $d\sigma/d\Omega$ & ** $^{\S\S\S)}$ \\
\hline
\end{tabular} 
\end{center}
\noindent
{$^{\dagger)}$: inconsistency with other data and model calculation at forward angles.} \\
{$^{\dagger\dagger)}$: inconsistency with other data and model calculation for 
$p_K \approx$ 530 MeV/c and for $\cos \theta \ge$ 0.5.} \\
{$^{\ddag)}$: shortcoming of model calculation at backward angles.} \\
{$^{\ddag\ddag)}$: inconsistency between the data.} \\
{$^{\S)}$: some inconsistencies between the data.} \\
{$^{\S\S)}$: ambiguities in the data analysis at forward angles.} \\
{$^{\S\S\S)}$: statistical fluctuations in the data.} \\
\end{table}

\ack{
A.S. acknowledges useful discussions with D. Bugg. 
This work was partially supported by Deutsche
Forschungsgemeinschaft through funds provided to the SFB/TR 16
``Subnuclear Structure of Matter''. This research is part of the \, EU
Integrated \, Infrastructure \, Initiative Hadron Physics Project under
contract number RII3-CT-2004-506078. A.S. acknowledges support by the
COSY FFE grant No. 41760632 (COSY-085) and the JLab grant SURA-06-C0452.}

%
\appendix
\section{ }
For convenience we collect here kinematical formulas
used for the transformation of the experimental results on
differential cross section from one system to another.
Let $A$ be a Lorentz invariant amplitude. The differential cross
section for the two-body reaction $a{+}b{\to}c{+}d$
in the laboratory frame is given as
\begin{eqnarray}
\frac{d\sigma}{d\Omega_{lab}}=\frac{p_c^2}{64\pi^2m_bp_a}\,\,
\frac{|A|^2}{(E_a+m_b)p_c-p_aE_c\cos\theta_{lab}} \ ,
\end{eqnarray}
where $p_a$, $E_a$, $p_c$, $E_c$ are the momenta and energies of the
particles $a$ and $c$ in the laboratory frame, while $\theta_{lab}$ is
scattering angle of particle $c$ with respect to the beam direction in 
laboratory. Here $m_a{=}m_c$ and $m_b{=}m_d$ are the masses of the
particles. The differential cross
section in the center of mass is given as
\begin{eqnarray}
\frac{d\sigma}{d\Omega_{cms}}=\frac{1}{64\pi^2s}|A|^2 \ ,
\end{eqnarray}
where the invariant collision energy squared is 
\begin{eqnarray}
s=m_a^2+m_b^2+2E_am_b \ .
\end{eqnarray}
The Lorentz invariant cross section is defined as
\begin{eqnarray}
\frac{d\sigma}{dt}=\frac{|A|^2}{64\pi sq_a^2} \ ,
\end{eqnarray}
where 
\begin{eqnarray}
q_a^2&=&\frac{(s-m_a^2-m_b^2)^2-4m_a^2m_b^2}{4s} \ , \nonumber \\
t&=&-2q_a^2(1-\cos\theta_{cms})=2m_a^2-2E_aE_c+2p_ap_c\cos\theta_{lab} \ .
\label{transc}
\end{eqnarray}
The relation between the scattering angles in center of mass and
laboratory frames is given as
\begin{eqnarray}
\tan\theta_{lab}=\frac{2m_b\sqrt{s}\, \sin\theta_{cms}}
{(s-m_a^2-m_b^2)\cos\theta_{cms}+s+m_a^2-m_b^2} \ .
\end{eqnarray}
Furthermore $E_c$ depends on the scattering angle and in more compact
form can be expressed in terms of invariants as
\begin{eqnarray}
E_c=\frac{s+t-m_a^2-m_b^2}{2m_b} \ ,
\end{eqnarray}
while $t$ is related to $\theta_{cms}$ and $\theta_{lab}$ by
Eq.~(\ref{transc}).

\end{document}